\documentclass[preprint,review,12pt]{elsarticle}
 
\usepackage[top=1in, bottom=1in, left=1in, right=1in]{geometry}

\usepackage{graphicx}
\usepackage{amsmath}

\usepackage{natbib}
\usepackage{amssymb}
\usepackage{amsbsy,amsmath}

\usepackage{amsthm}

\usepackage{lineno}
\usepackage{subfig}
\usepackage{enumerate}
\usepackage{fullpage}
\usepackage{algorithm}
\usepackage{algpseudocode}
\usepackage[colorinlistoftodos]{todonotes}

\usepackage{mathtools}

\newcommand{\bs}[1]{\boldsymbol{#1}}
\usepackage{hyperref}

\biboptions{sort,compress} 

\usepackage{xcolor}

\newcommand*\patchAmsMathEnvironmentForLineno[1]{%
  \expandafter\let\csname old#1\expandafter\endcsname\csname #1\endcsname
  \expandafter\let\csname oldend#1\expandafter\endcsname\csname end#1\endcsname
  \renewenvironment{#1}%
     {\linenomath\csname old#1\endcsname}%
     {\csname oldend#1\endcsname\endlinenomath}}%
\newcommand*\patchBothAmsMathEnvironmentsForLineno[1]{%
  \patchAmsMathEnvironmentForLineno{#1}%
  \patchAmsMathEnvironmentForLineno{#1*}}%
\AtBeginDocument{%
\patchBothAmsMathEnvironmentsForLineno{equation}%
\patchBothAmsMathEnvironmentsForLineno{align}%
\patchBothAmsMathEnvironmentsForLineno{flalign}%
\patchBothAmsMathEnvironmentsForLineno{alignat}%
\patchBothAmsMathEnvironmentsForLineno{gather}%
\patchBothAmsMathEnvironmentsForLineno{multline}%
}

\graphicspath{ {./figs} }
\journal{Elsevier} 

\begin{document}
\begin{frontmatter}

\title{Data-Driven CFD Modeling of Turbulent Flows Through Complex Structures}


\author{Jian-Xun Wang\corref{corjxw}}
\author{Heng Xiao\corref{corxh}}
\cortext[corxh]{Corresponding author. Tel: +1 540 231 0926}
\ead{hengxiao@vt.edu}

\address{Department of Aerospace and Ocean Engineering, Virginia Tech, Blacksburg, VA 24060, United States}

\begin{abstract} 
  The growth of computational resources in the past decades has expanded the application of
  Computational Fluid Dynamics (CFD) from the traditional fields of aerodynamics and hydrodynamics
  to a number of new areas. Examples range from the heat and fluid flows in nuclear reactor vessels
  and in data centers to the turbulence flows through wind turbine farms and coastal vegetation
  plants. However, in these new applications complex structures are often exist (e.g., rod bundles
  in reactor vessels and turbines in wind farms), which makes fully resolved, first-principle based
  CFD modeling prohibitively expensive. This obstacle seriously impairs the predictive capability of
  CFD models in these applications.  On the other hand, a limited amount of measurement data is
  often available in the systems in the above-mentioned applications. In this work we propose a
  data-driven, physics-based approach to perform full field inversion on the effects of the complex
  structures on the flow. This is achieved by assimilating observation data and numerical model
  prediction in an iterative Ensemble Kalman method. Based on the inversion results, the velocity
  and turbulence of the flow field can be obtained. A major novelty of the present contribution is
  the non-parametric, full field inversion approach adopted, which is in contrast to the inference
  of coefficient in the ad hoc models often practiced in previous works. The merits of the proposed
  approach are demonstrated on the flow past a porous disk by using both synthetic data and real
  experimental measurements. The spatially varying drag forces of the porous disk on the flow are
  inferred. The proposed approach has the potential to be used in the monitoring of complex system
  in the above mentioned applications.
\end{abstract}

\end{frontmatter}


\section{Introduction}
\label{sec:intro} 


In the past several decades, the growth of computational resources along with the theoretical and
algorithmic development in computational fluid dynamics (CFD) has led to a large number of
commercial and open-source codes in the fields ~\cite{weller1998tensorial}. These powerful software
tools have enabled scientists and engineers to numerically simulate the heat and fluid flows in
complex systems at unprecedented fidelities.  These numerical simulations have played critical roles
in scientific enquiry, engineering design and decision making, and operational forecasting involving
complex systems. For example, in commercial aircraft design, CFD simulations have largely replaced
wind tunnel testing, reducing the required number of wind tunnel tests from 77 in the 1980s to less
than ten in present day.  Consequently, the use of CFD tools has led to drastic reduction of the
development cycle duration and costs~\cite{johnson2005thirty,johnson1992tranair}.

The ever-increasing availability of computational resources has also expanded the application of CFD
from the traditional fields of aerodynamics and hydrodynamics to a number of new areas, where the
design, optimization, and other decision processes has traditionally been supported by using
empirical models. Examples include simulations of the mass, heat and fluid flows in nuclear reactor
vessels, the flow in wind turbine farms, the flow and convective heat transfer in data centers, and
the study of sediment and turbulence in the flow through coastal vegetation plants, just to name a
few.  In the nuclear energy industry, CFD has been increasingly used for safety assessment of the
reactor vessels in nuclear power plants~\cite{bieder03,iaea02,hoehne04,Bertolotto:09cy}; in coastal
engineering, numerical models have enabled forecasting of flood inundations, which provided valuable
support for decision-making in emergence management~\cite{luettich1992adcirc}; CFD has also been
used to aid heat management of data centers by modeling the heat generated by computer racks and the
convective heat transfer in the rooms~\cite{hamann2008measurement, fernando2012can}.

It is well known that successful predictions of the simulations rely on faithful representations of
the system geometry, initial and boundary conditions, materials properties, and the important
physics in the numerical model. However, many systems intrinsically exhibit multi-scale features
that prevents first-principle based representation in numerical simulations or makes such
representations very expensive. For example, a first-principle based simulation of the flow field in
a wind turbine farm would require detailed representation of the rotating blades of the turbines,
whose geometric features have a much smaller length scale ($\sim 0.1$~m) than that of the wind farm
(several kilometers).  Similarly, a full fidelity simulation of the currents and waves in a coastal
region covered with vegetation plants requires resolution of the plant geometry, which not only have
complex geometries and small-scale features (e.g., stems and leafs) but are also flexible. That is,
the plants deform in response to hydrodynamic forces and the effects on the flow may change
accordingly. In nuclear reactor vessels, the coolant passes through thin channels in the rod bundle
(fuel assembly), and the sizes of the channels are many orders of magnitude smaller than that of the
reactor vessel.  Because of the prohibitive computational costs of the first-principle based
simulations, numerical modelings of such problems have inevitably relied on ad hoc
parameterizations, which are often developed based on physics-based reasoning but with drastic
simplifications. In the examples above, the wind turbines are often represented as actuation
disks~\cite{ammara2002viscous,wu2011large}, the vegetation plants are modeled as
macro-roughness~\cite{masterman92predict,Doncker2009}, and the rod bundles in reactor vessels are
often modeled as porous media with resistance. All these ad hoc models are essentially momentum sink
with their resistances to the flow correlated to the local flow velocity and the physical properties
of the structure (turbines or plants). The same concept can be extended to the heat and fluid flows
in data centers, where the computer racks can be described as volumetric heat sources without
explicitly representing their detailed geometries.  While these parameterization techniques have
led to significant reduction of computational costs compared to first-principle based modeling,
they also introduce large model uncertainties in the numerical predictions due to their ad hoc
nature.

Fortunately, the increasing availability of observation data and recent development of data
assimilation algorithms~\cite{evensen2003ensemble, pocock2011state, iglesias2013ensemble} has opened
new possibilities for computational mechanics simulations with parameterized models. The wide spread
deployment of measurement instrument made it possible to integrate data with numerical models in a
data-driven modeling approach. For example, this data-driven approach has been explored in the
modeling of wind turbine farms~\cite{iungo2015data} and in quantifying uncertainty in turbulence
models~\cite{xiao-mfu, wu2015bayesian}.  By assimilating the available observation data, the uncertain model
parameters can be inferred based on the data assimilation techniques.  Although the inverse modeling
has been recently used in the CFD literature to infer the coefficients in turbulence
models~\cite{kato2013approach, kato2015data}, an important limitation of the parametric approach is
that it is still constrained by the basic assumptions of the ad hoc model and thus does not fully
explore the space of model uncertainties. That is, it is possible that truth cannot be described by
the chosen model regardless of how the coefficients are varied.  For example, when modeling wind
turbines with CFD solvers based on actuation disk model, the drag of the turbine on the flow is
often made proportional to the square of the velocity in the immediate upstream of the
turbine~\cite{wu2011large}. The parametric approach can only calibrate the proportionality constant
but does not alter the form of the ad hoc model. To overcome this limitation, in this work we
propose a non-parametric, full-field approach and use inverse modeling to infer the optimal spatial
force distribution to represent the effects of complex structures based on observation data.  The
non-parametric approach is able to explore a much larger uncertainty space (i.e., the spatial
distribution of the forces in the example of wind turbines and vegetation) than in the parametric
approach.  The proposed inverse modeling method is most valuable in two scenarios where observation
data are available and utilized, i.e., (1) forecasting of complex systems and (2) calibrating
numerical models and guiding the model development.  CFD simulations are often used along with
monitoring data to provide \emph{forecasting} of complex systems.  This technique, referred to as
data assimilation, has long been used in operational weather
forecasting~\cite{kalnay2003atmospheric}, but its potential has also been increasingly realized in
the CFD community~\cite{kato2013approach, kato2015data} and beyond. 
For example, the US Air Force has developed a plan to create a Digital Twin for every aircraft platform, 
which aims to predict the damage initiation and accumulation throughout its service life and 
thus provide support for decision-making in fleet management~\cite{smarslok2012error}.  
Observation data can be used to infer and calibrate the parameterization of the complex 
structures (e.g., wind turbines or vegetation), which can lead to \emph{improvement} 
thereof and provide \emph{guidance} for the model development.


In this work we use the flow past a porous disk as shown in Fig.~\ref{fig:exp1} to illustrate
the potential of the proposed data-driven and inverse modeling approach based on simulations with
reduced order representation of the porous disk.  Specifically, we performed experimental
measurements and performed CFD simulations where the disks are represented with spatially varying
drag force field, and then the experimental data are used to infer the optimal representation of the
porous disk in the CFD simulations. The novelty lies in the non-parametric, full-field
representation of the porous disks, which has implication for using computational mechanics
simulations in a wide range of real-word engineering applications from nuclear power plants and wind
farm to vegetation plants modeling and data centers.

The rest of the paper is organized as follows. The formulation and experimental data of the example
problem, the flow past a porous disk, is presented in Section~\ref{sec:prob}, and the data-driven
inverse modeling approach is introduced in Section~\ref{sec:meth}.  Numerical results of the example
problem are presented in Section~\ref{sec:result} to assess the merits and limitation of the
proposed method. The significance and limitations of the proposed inverse modeling are 
discussed in Section~\ref{sec:dis}. Finally, Section~\ref{sec:con} concludes the paper.

\section{Problem Background, Formulation and Experimental Data}
\label{sec:prob} 
In the example problem, we aim to predict the flow in the wake of a porous disk, which is often
used to represent turbines (as well as propulsion devices) in lab experiments~\cite{myers2010experimental}. 
Studies on the wake of an energy-harvesting structure and its interactions with 
the downstream devices and the atmospheric or seabed boundary 
layers are of critical importance. It is because a better understanding of the corresponding 
flow field assists the optimization of wind turbine layout and helps to assess the environmental 
footprint of energy harvesting projects~\cite{sorensen2002numerical, ammara2002viscous, 
vermeer2003wind, gomez2005anisotropy, gebreslassie2012cfd, gebreslassie2013-1, gebreslassie2013-2}. 
A number of investigators have performed large eddy 
simulations to advance the understanding of the wake structure~\cite{churchfield2013large, 
calaf2010large, troldborg2011numerical, porte2011large, wu2011large, yang2012computational}.
However, as mentioned above, first-principle based simulations with full resolution of the turbine
blade geometry are computationally expensive, and one often has to resort to lower fidelity models
based on reduced order representation of the turbines, particularly in operational forecasting,
where a positive lead-time is required. For example, computational cost can be greatly reduced by
using Reynolds-Averaged Navier--Stokes (RANS) equations to model the fluid flow and actuation disk
model to represent the hydrodynamic effects of the turbines. With the actuation disk model, 
explicit meshing of the turbine geometry is avoided. The momentum equation of the flow field
reads as follows:
\begin{align}
 \frac{\partial U_i}{\partial t}+\frac{\partial U_i U_j }{\partial x_j} = & -
\frac{1}{\rho} \frac{\partial p}{\partial x_i} +\nu\frac{\partial^2 U_i}{\partial x_j \partial x_j}
- \frac{\partial \tau_{ij}}{\partial x_j} + f_i,
  \label{eq:common-u} 
\end{align}
where $t$ and $x_i$ are time and space coordinates, respectively; $\rho$ and $\nu$ are the density
and viscosity of water, respectively; $U_i$ and $p$ represent Reynolds-averaged velocity and
pressure, respectively.  {\color{black} The body force term $f_i$ is a momentum sink used to account
  for the hydrodynamic effects of the turbines, which is often computed from actuation disk
  models~\cite{sorensen2002numerical} and is active only in the regions of the computational domain
  that are occupied by the turbines.  Finally, $\tau_{ij}$ is the Reynolds stresses computed by
  using an eddy viscosity model:
\begin{equation}
 \tau_{ij} = - \frac{2}{3}k \delta_{ij} + \nu_t\left( \frac{\partial U_i}{\partial
    x_j} +\frac{\partial U_j}{\partial x_i} \right),
\end{equation}
where $\delta_{ij}$ is the Kronecker delta, $k$ is the turbulent kinetic energy (TKE), and $\nu_t$ is 
the turbulent eddy viscosity modeled by the standard two-equation $k$--$\varepsilon$
model~\cite{wilcox98}:
\begin{align}
	\label{eq:k}
	\frac{\partial k}{\partial t} + U_j \frac{\partial k}{\partial x_j} = &
	P_k - \varepsilon 
	+ \frac{\partial }{\partial x_j} \left[ (\nu + \nu_t/\sigma_k)
	\frac{\partial k}{\partial x_j}
	\right] \\
	\label{eq:eps}
	\frac{\partial \varepsilon}{\partial t} + U_j \frac{\partial \varepsilon}{\partial x_j} = &
        C_1 \frac{\varepsilon  P_k}{k} - C_2 \frac{\varepsilon^2}{k} + \frac{\partial }{\partial x_j}
        \left[ (\nu + \nu_t/\sigma_\varepsilon)
          \frac{\partial \varepsilon}{\partial x_j} \right] \\
        & \textrm{with } \quad P_k = \tau_{ij} \frac{\partial U_i}{\partial x_j}
\end{align}
where $\varepsilon$ is dissipation rate,
\begin{equation*}
	\label{eq:nut}
	\nu_t = C_\mu k^2/\varepsilon ,
\end{equation*}
 and the coefficients are:
\begin{equation*}
	C_1 = 1.44,\;  C_2 = 1.92, \; C_\mu = 0.09, \; \sigma_k = 1.0, \;
	\sigma_\varepsilon = 1.3.
\end{equation*}
Note that the term $P_k$ in Eq.~(\ref{eq:k}) denotes the production of TKE by
extracting the kinetic energy from the mean flow, which relies on the presence 
of the mean velocity gradient. Similarly, the term $C_1 \frac{\varepsilon  P_k}{k}$ 
in Eq.~(\ref{eq:eps}) is the production term for the dissipation rate $\varepsilon$. 
}

\begin{figure}[!h]
\begin{center} 
\subfloat[Overview of the water channel]{\includegraphics[width=0.6\textwidth]{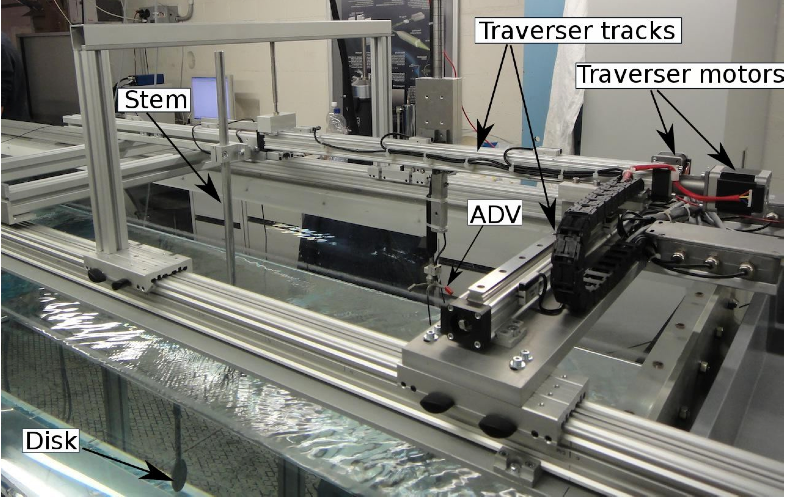}}\\
\subfloat[The porous disk]{\includegraphics[width=0.6\textwidth]{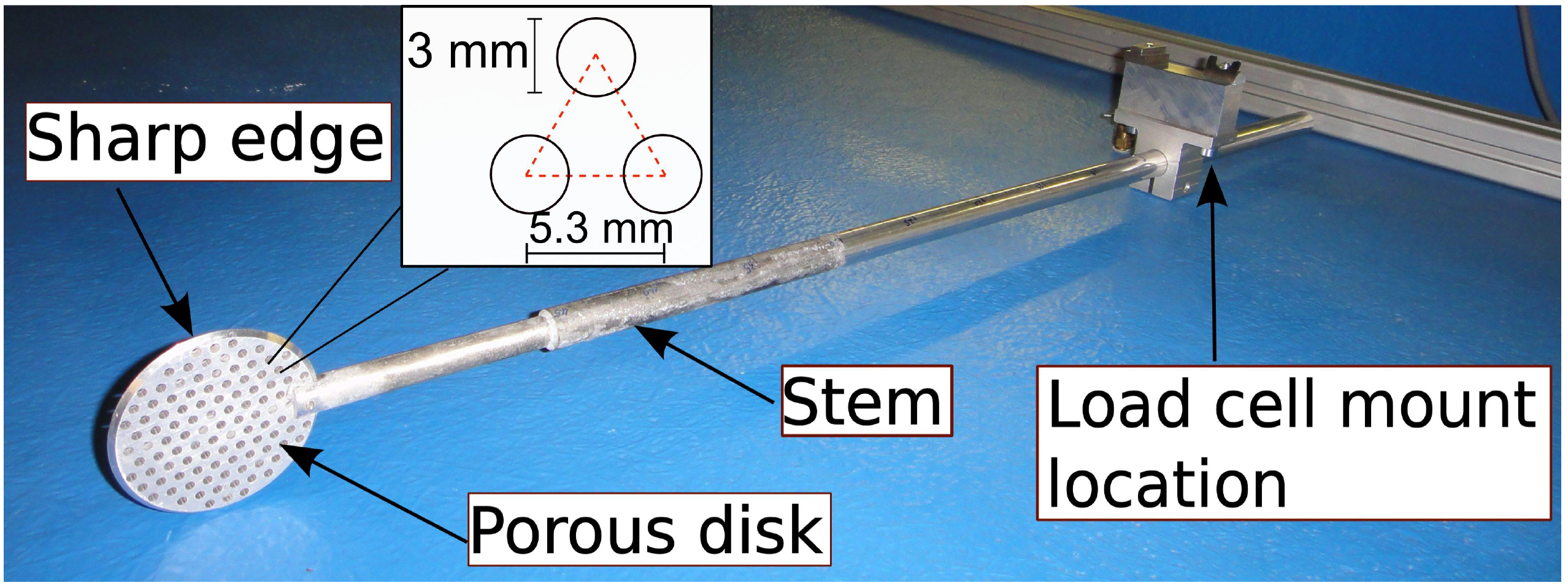}}
\end{center}
 \caption{(a): Overview of the water channel with the porous disk, the stem, the ADV, and
 the traverser system annotated with labels. The nominal water depth is 40 cm. 
 (b) A porous disk used in this study as a simple representation of a turbine~\cite{xiaoexperimental}.}
\label{fig:exp1}
\end{figure}

In this work, the same set of equations and actuation disk models outlined above are used to describe
the flow and the porous disk. Unlike computing the drag force $f_i$ with an assumed force
distribution in the traditional CFD modeling approaches, a number of measurements of the flow field
(e.g., velocity) are incorporated in the proposed data-driven approach to infer an optimal force
distribution in the region occupied by the disk. Therefore, we have conducted a laboratory
experiment to measure the flow velocity and TKE in the wake of the
disk. The experiment was conducted in a recirculating channel in the Institute of Fluid Dynamics at
ETH Zurich (see ref.~\cite{xiaoexperimental} for details).  The dimension of test section of this
channel is 2.3 m (length) $\times$ 0.45 m (width) $\times$ 0.4 m (water depth). A single porous disk
is mounted on a stem inside the water channel (Fig.~\ref{fig:exp1}a). The disk is manufactured with
a diameter of $D = 92$ mm and thickness of $h = 4$ mm. The holes on the disk have a diameter of $3$
mm and are separated by a distance of 5.3 mm (Fig.~\ref{fig:exp1}b). The disk is arranged coaxially
in the flow direction with its center positioned at $2.25D$ (227~mm) above the bottom of the
channel, and thus the interference of the disk with the bottom boundary layer of the channel is
negligible, at least in the near-wake region studied in this work. Therefore, the mean flow can be
considered axisymmetric. An acoustic Doppler velocimeter (ADV) installed on a traversing system is
used to measure velocities in the wake of the disk. The detailed experimental setup is shown in
Fig.~\ref{fig:exp1}a.

\begin{figure}[htbp]
\begin{center}
\includegraphics[width=0.7\textwidth]{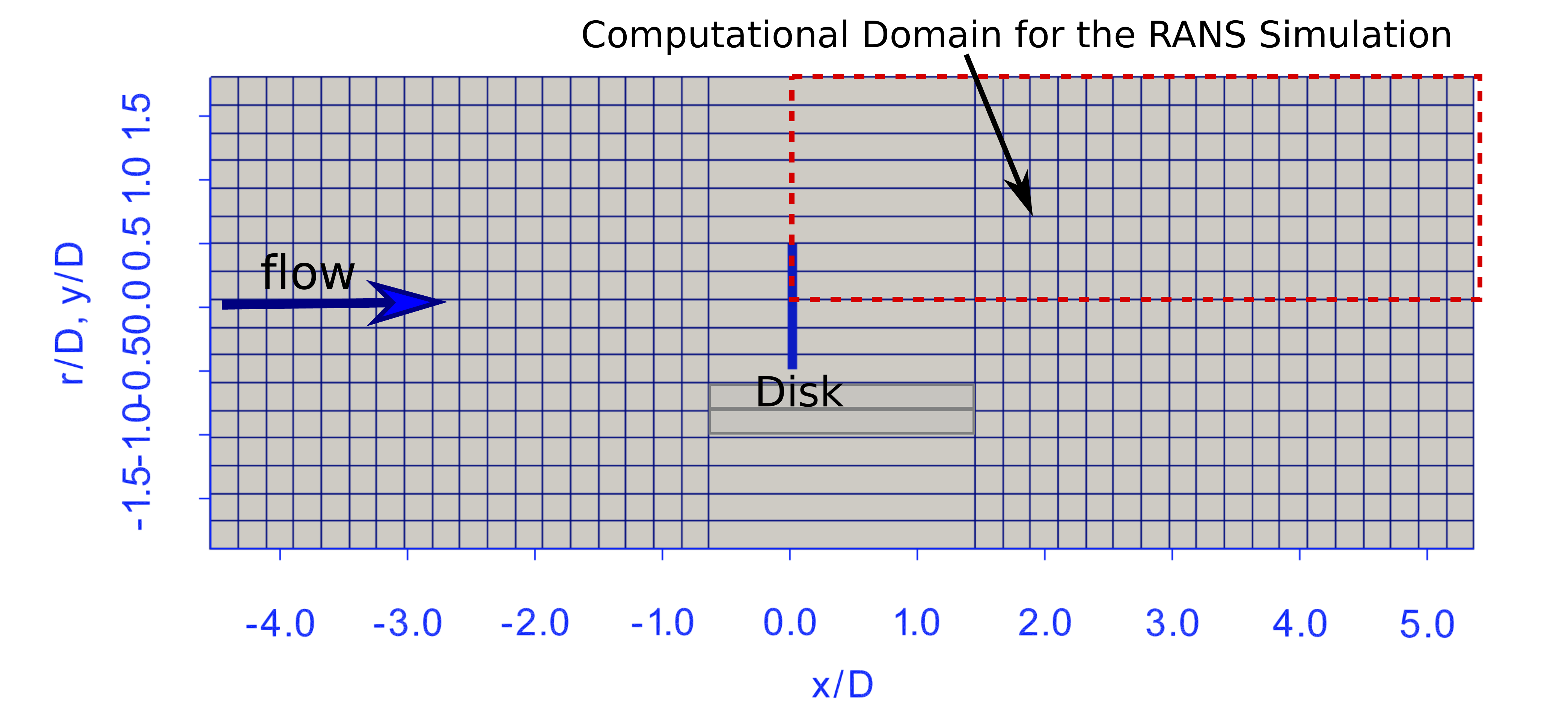}
\caption{Locations of the velocity measurements upstream and downstream from Disk, which is
indicated by a thin strip. The shown streamwise and radial coordinates are normalized by disk
diameter D and the origin is located at the disk center.}
\label{fig:exp2}
\end{center}
\end{figure}

The measurements used as observations for the inverse modeling are conducted at 
the disk-center height in a single horizontal plane, which are on a grid of $38 \times 18$ 
points with intervals of $2$ cm $\times$ $2$ cm, or $0.22D \times 0.22D$, 
in streamwise and spanwise directions. The measurement locations are illustrated in 
Fig.~\ref{fig:exp2}, with $18 \times 19$ sets of velocities and TKEs measured 
in the downstream domain. Considering the symmetry of the disk, only half
of that domain needs to be simulated. The spatial averaged measurements, located in the
region surrounded with the red dashed line, are used as observation data for the inversion.  
The total force $f_t$ acted on the disk is also measured by using a six-axis load
cell mounted on the stem (see Fig.~\ref{fig:exp1}b). The measured total force is used
as the input in the proposed data-driven approach, and the objective is to infer how 
to distribute the force $f_t$ over the disk.  


\section{Data-Driven Inverse Modeling Approach}
\label{sec:meth}
In this work we proposed a data-driven, inverse modeling approach for modeling the
flow with complex structures. This approach consists of two parts: reduced order
representation of the complex structures (i.e., porous disk in the example problem),
and the ensemble Kalman method based inverse modeling~\cite{iglesias2013ensemble}. 
Specifically, the momentum sink due to the disk is represented by a non-parametric, full-field
approach. Orthogonal basis functions are utilized to reduce order of the 
full-field force representation. The corresponding coefficients of these
basis functions are inferred by incorporating the measurements of the flow 
quantities (e.g., velocity and turbulence kinetic energy) based on an ensemble
based Kalman method. The details of the two aspects are presented below.

\subsection{Reduced Order Representation of Momentum Sink Due to Disk}
\label{sec:forward}

Here we only consider the drag force acting by the disk on the flow in the streamwise
direction, and the force is indicated as $f$ hereafter without ambiguity.  In actuation
disk theory, the drag forces are often assumed to be uniform over the disk or proportional 
to the amount of velocity drop across the disk.  Another approach is to compute the force 
distribution with potential flow theory by assuming that the porosity of the disk 
is zero~\cite{currie2012fundamental}.

In the present approach, we allow the body force~$f(x)$ to have any physically reasonable 
spatial field constrained only by the axial symmetry with respect to the center of the disk. 
Furthermore, we represent the field with a number of orthogonal basis functions~$\phi_\alpha$, 
i.e., 
\begin{align}
  \label{eq:para} 	
  f(x) & = f_t \Phi(x)   \\ 
  \mathrm{ with} \quad \Phi(x)  & = \sum_{\alpha = 1}^{M} w_\alpha \phi_\alpha (x) \notag
\end{align}
where $f_t = \int_\Omega f(x) dx$ is the magnitude of  drag force in the 
computational domain~$\Omega$, $\Phi(x)$
is the distribution function in the domain with a normalization condition~$\int_{\Omega}
\Phi(x) d\Omega = 1$, and $M$ is the truncated number of basis functions used to 
represent the force distribution. Since it is more convenient to describe the geometry 
of circular disk in polar coordinate~$x = (r, \theta)$, the orthogonality of the basis 
functions~$\phi_{\alpha}(x)$ should be ensured in polar coordinate as,   
\begin{equation} 
\int_{0}^{2\pi}\int_{0}^{0.5D}{\phi_{\alpha}} \phi_{\beta}rdrd\theta = \delta_{\alpha \beta},  
\quad 
\textrm{with} \;
\alpha, \beta = 1, 2, \cdots, M
\end{equation} 
in which $D$ is the diameter of the disk and $\delta_{\alpha \beta}$ denotes Kronecker delta.
Commonly used basis functions (e.g., Chebyshev polynomials, Fourier series) have orthogonality only
in Cartesian coordinates and not in polar coordinate. Therefore, we choose a set of orthogonal basis
functions $\phi_{\alpha}(r, \theta)$ in polar coordinate introduced in
ref.~\cite{verkley1997spectral}, which are defined as
\begin{align}
\phi_{\alpha}(r, \theta) = W_{m, \alpha}(r)e^{im\theta}, 
\label{eq:common-p}
\end{align}
where $i$ is an imaginary unit and
\begin{equation}
  \label{eq:nut} 
  W_{m,\alpha}(r) = r^{|m|}P^{(0, |m|)}_{c}(s),
\end{equation}
and $P_c^{(0, |m|)}(s)$ is a Jacobi polynomial with argument $s = 2r^2 -1$ and degree $c = - 0.5
|m|$, which are orthogonal with respect to the weight~$(1 - s)^0 (1 + s)^{|m|}$; the integer~$m$ can
be chosen from $-\infty$ to $\infty$, which is set as $m = 0$ in this work. The first three basis
functions are $\phi_1 = 2r^2-1$, $\phi_2 = 6r^4 - 6r^2 + 1$, and $\phi_3 = 20r^6 - 30r^4 + 12r^2-1$.

\subsection{Inverse Modeling Based on Ensemble Kalman Method}
\label{sec:enkf} 
With the set of orthogonal basis functions introduced above, the momentum sink due to the          
porous disk can be represented by $M$ coefficients 
$\bs{\omega} = [\omega_1, \cdots, \omega_{M}]^{T}$ corresponding to these basis functions.  
When some sparse measurements of the flow field (e.g., velocity, turbulence kinetic energy) 
are available, these unknown coefficients can be inferred based on the Bayesian inference 
approach. In this work an iterative, ensemble-based Bayesian inference 
method~\cite{iglesias2013ensemble} is employed to perform the inversion. This technique 
is closely related to the ensemble-based filtering methods (e.g., ensemble Kalman filter), 
in which the statistical mean and covariance are estimated based on the 
samples~\cite{evensen1994sequential}. An overview of the ensemble Kalman method 
based inverse modeling procedure is presented in Fig.~\ref{fig:enkfFlow}. 
To infer the distribution of the force due to porous disk, the corresponding unknown 
coefficients $\bs{\omega}$ are augmented to the physical state (i.e., velocity field $\bs{u}$ 
and turbulence kinetic energy field $\bs{k}$). The augmented state vector is denoted as 
$\bs{x} \equiv [\bs{u}, \bs{k}, \bs{\omega}]^{T}$, and the ensemble of the state is denoted
as $\{\bs{x_j}\}_{j = 1}^{N}$, where $N$ is the number of samples. Given the observation data
of the flow velocity $\bs{u_{o}}$ and turbulence kinetic energy $\bs{k_o}$, the inversion 
proceeds as follows:
  
\begin{figure}
\begin{center}
\noindent
\includegraphics[width=24pc]{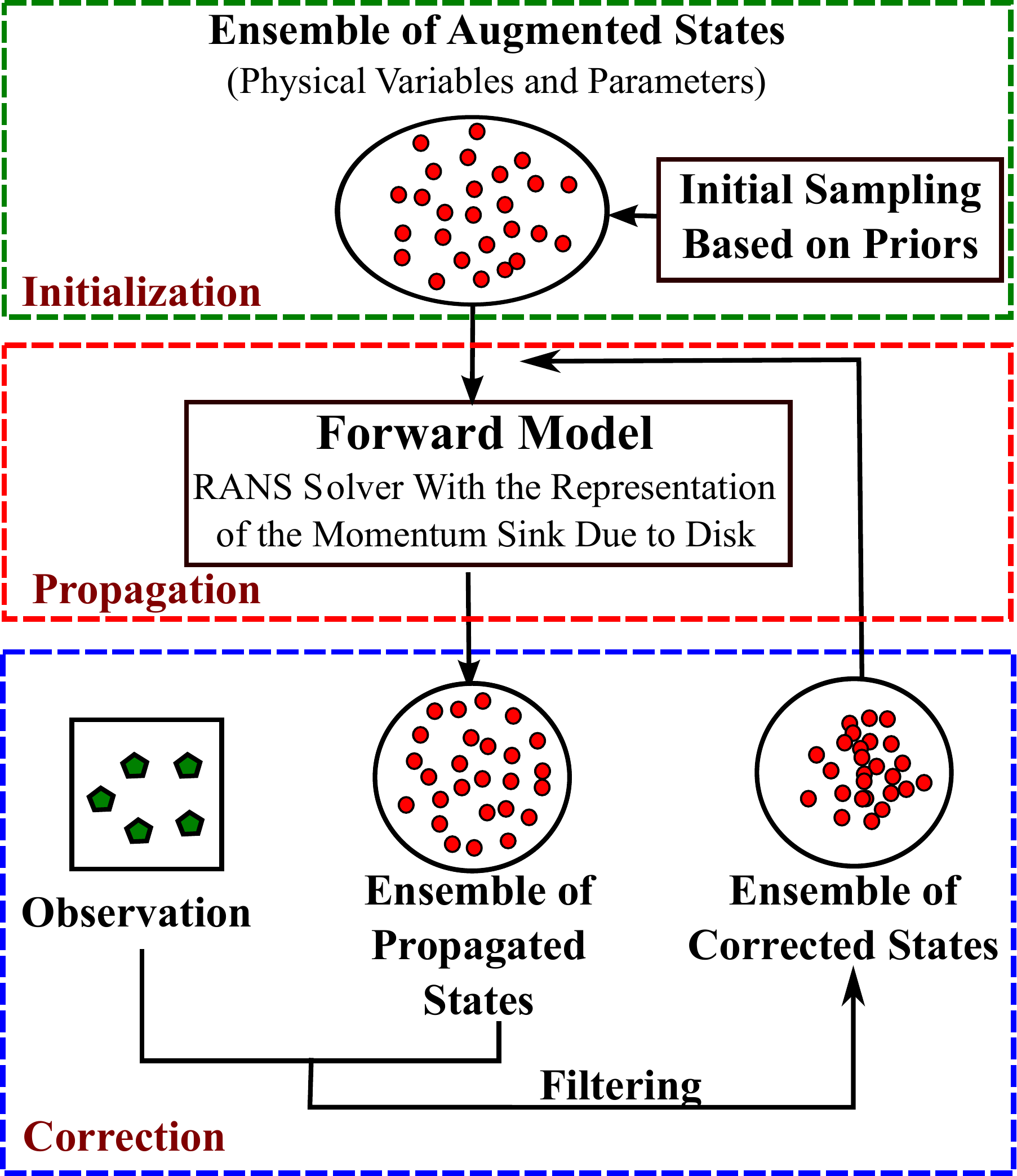}
\caption{Schematic of the inverse modeling approach based on the state augmentation.  The system
  state is augmented to include both the physical state (velocity and turbulence kinetic energy) and the
  parameters to be inferred (coefficient vector $\bs{\omega}$ for disk representation). An ensemble 
  representing the augmented state is propagated via the forward model.  
  The propagated ensemble is then updated in the correction process 
  based on the observation data.  The updated state (physical
  quantities and model parameters) is set as new initial condition, and this iteration continues 
  until the statistical convergence is reached.}
\label{fig:enkfFlow}
\end{center}
\end{figure}

\begin{enumerate}
\item \textbf{Sampling of prior distribution.} 
First, one should provide an initial guess of how the force is distributed, which can be 
based on the physical knowledge or the low-fidelity models, e.g., 
actuation disk model. This initial guess is then perturbed based on one's confidence on
it. These perturbed force distributions are known as the prior ensemble. That is, given 
the prior for force distribution coefficients $\bs{\omega}$, $N$ samples of $\bs{\omega}$ 
are drawn. Each sample represents a possible force distribution.

\item \textbf{Propagation.} In this step the flow field corresponding to each
sample of the force distributions is simulated by solving the forward 
model (i.e., RANS equations). The physical states (i.e., velocity and turbulence
kinetic energy) are predicted, and the propagated ensemble $\{\hat{\bs{x}}_j\}_{j = 1}^M$
is obtained. The mean $\bar{\bs{x}}$ and covariance $P$ of the propagated ensemble 
are estimated. Note that enough forward time steps are required to ensure that the 
steady state is achieved in each propagation step.
 
\item \textbf{Correction.} The propagated velocity and turbulence kinetic energy
are compared with the observation data. Note that the dimension of observed 
state vector~$\bs{y}_o$ is much smaller than that of the full state~$\bs{x}$, 
since the flow fields can only be measured sparsely in most cases due to the intrusive
nature of measurement techniques (e.g., Acoustic Doppler velocimetry). 
To perform the correction, the model predictions and the measurements are linearly
combined. In the combination the weight of each component is determined by the
Kalman gain matrix $K$, which is computed with the ensemble covariance $P$
and observation error covariance $R$. After correction, the analyzed state 
contains the updated parameters of force distribution. 

\end{enumerate}

The propagation and correction steps (step 2 and 3) are repeated until the ensemble is statistically
converged. The convergence is achieved when the variance of ensemble scattering is much
smaller than the variance of the observation noise. The algorithm of the inversion scheme
is summarized in~\ref{app:enkf}, and further detail can be found 
in Ref.~\cite{iglesias2013ensemble}.

\section{Numerical Results}
\label{sec:result}

\subsection{Verification with Synthetic Data}
\label{sec:result1}
To establish confidence of the proposed inverse modeling approach, we
conduct a synthetic case to assess its performance. In this case the observations 
are generated by the forward model with given ``truth" of the inferred quantities, 
i.e., the force distribution. This specified force distribution is called as the 
synthetic truth, which is unknown in the inversion process. We sparsely select the 
velocities and turbulence kinetic energy (TKE) simulated with the synthetic truth as 
the synthetic observations by adding some white noise. The reason of
using synthetic data (i.e., specified spatial force distribution) to verify the
proposed method is that the true distribution is not available in the experiment.
Only the total force acted on the disk has been measured. 
On the other hand, even if the truth of force distribution is known, it is still difficult
to directly assess the performance of the inversion due to inadequacy of the 
forward model (e.g., the inadequacy in turbulence model), 
which poses difficulties to differentiate the errors caused by the inversion procedure 
and those due to the forward model.

In this work, we specify the synthetic true force distribution to be the potential flow solution
by assuming that porosity of the disk is zero~\cite{currie2012fundamental}. 
Consequently, the streamwise drag force per unit area can be expressed as,
\begin{equation}
f(r) = f_t  \lambda (1 + \tilde{\omega} r^2),  \quad \textrm{with} \, 0 \le r \le 0.5D,
\end{equation}
where $\tilde{\omega} = -4.0$, and $\lambda$ is a normalization factor.       
The synthetic observation data are generated by running the forward
model (i.e., RANS solver with the momentum sink) with this specified 
force distribution.
Specifically, we choose 100 observation locations that are uniformly distributed at 
each of the eight spanwise lines (i.e., $x/D = 0.01, 0.1, 0.5, 1.0, 1.5, 2.0, 2.5$, and $3.0$). 
The simulated velocities and TKE on these locations are observed, and zero mean 
Gaussian noises ($\epsilon_o \sim \mathcal{N}(0, \sigma_o^2)$) are added as the 
observation error, where the standard deviation $\sigma_o$ is $1\%$ of the truth. 

The aim here is to infer the parameter $\tilde{\omega}$ by using the proposed inversion scheme based
on the synthetic observation data. By comparing the inferred result with the synthetic truth
$\tilde{\omega} = -4.0$, the performance of inversion scheme can be evaluated. In this verification
case we consider two scenarios of increasing difficulty levels. For the first scenario, the initial
guesses of $\tilde{\omega}$ are uniformly drawn from the interval of $[-8, 0]$, whose mean equals to
the synthetic truth, i.e., $\mathop{\mathbb{E}}(\tilde{\omega}) = -4.0$, where
$\mathop{\mathbb{E}}(\cdot)$ denotes expectation. That is, the prior estimation of the force
distribution is unbiased. In the second scenario, a larger sampling interval of $[-14, 0]$ with a
biased mean $\mathop{\mathbb{E}}(\tilde{\omega}) = -7.0$ is applied. Uniformly distributed prior is
representative of lack of knowledge on the force distribution to be inferred in realistic problems.
For CFD problems, the ensemble size dominates the computational costs of the inversion
procedure. This is because for each sample of the force distribution, a model evaluation of RANS
equation is needed to obtain the corresponding flow field, which is used to compare with the
observation data in the correction step. We have studied the effects of ensemble size $N$ on the
inversion results for this problem and found that the results are not sensitive to the ensemble size
when $N$ is larger than eight. Therefore, a relatively small ensemble size of ten is adopted in this
study to reduce the computational cost.

\begin{figure}[!h]
	\begin{center} 
		\subfloat[With unbiased prior]{\includegraphics[width=0.5\textwidth]{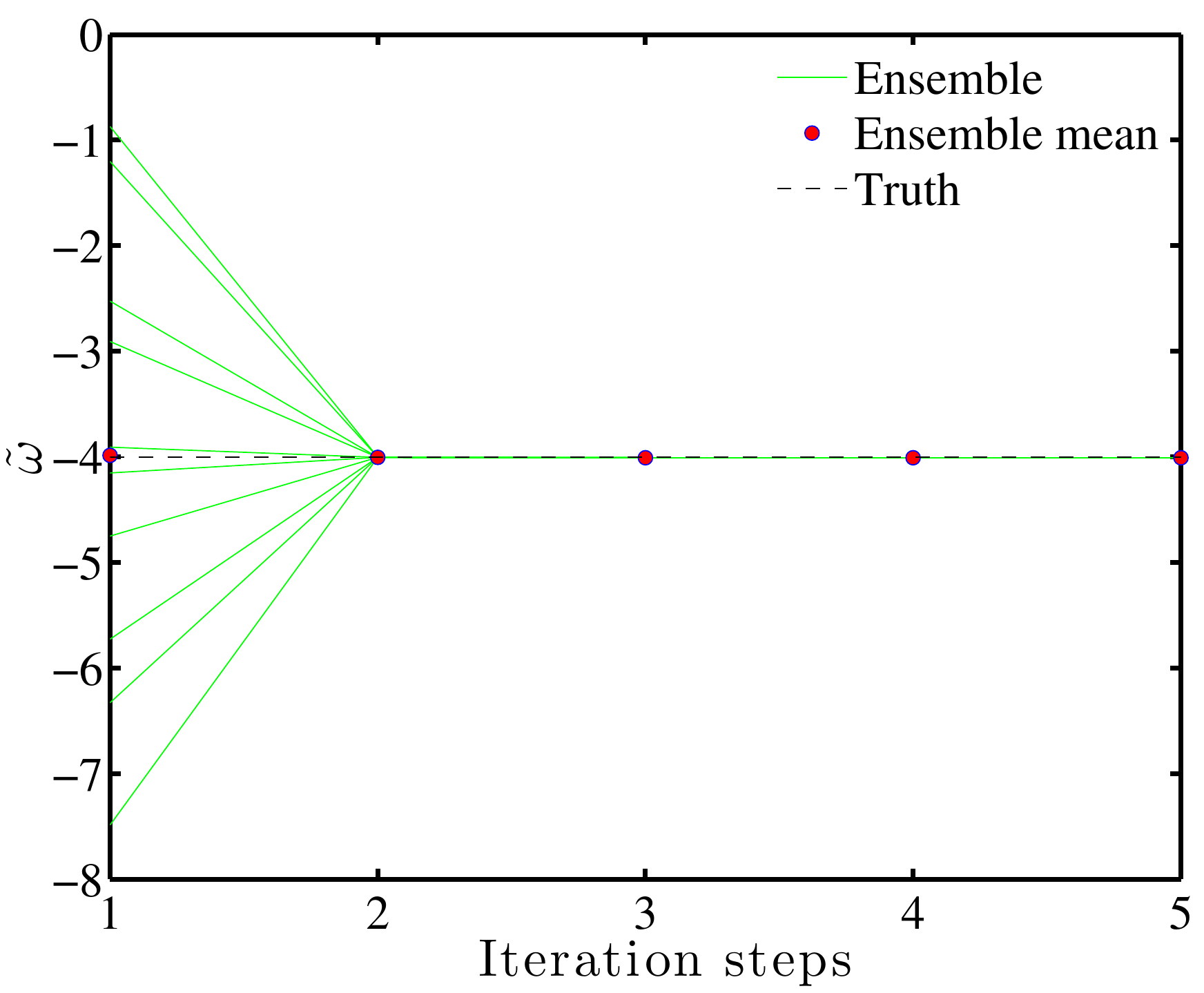}}
		\subfloat[With biased prior]{\includegraphics[width=0.5\textwidth]{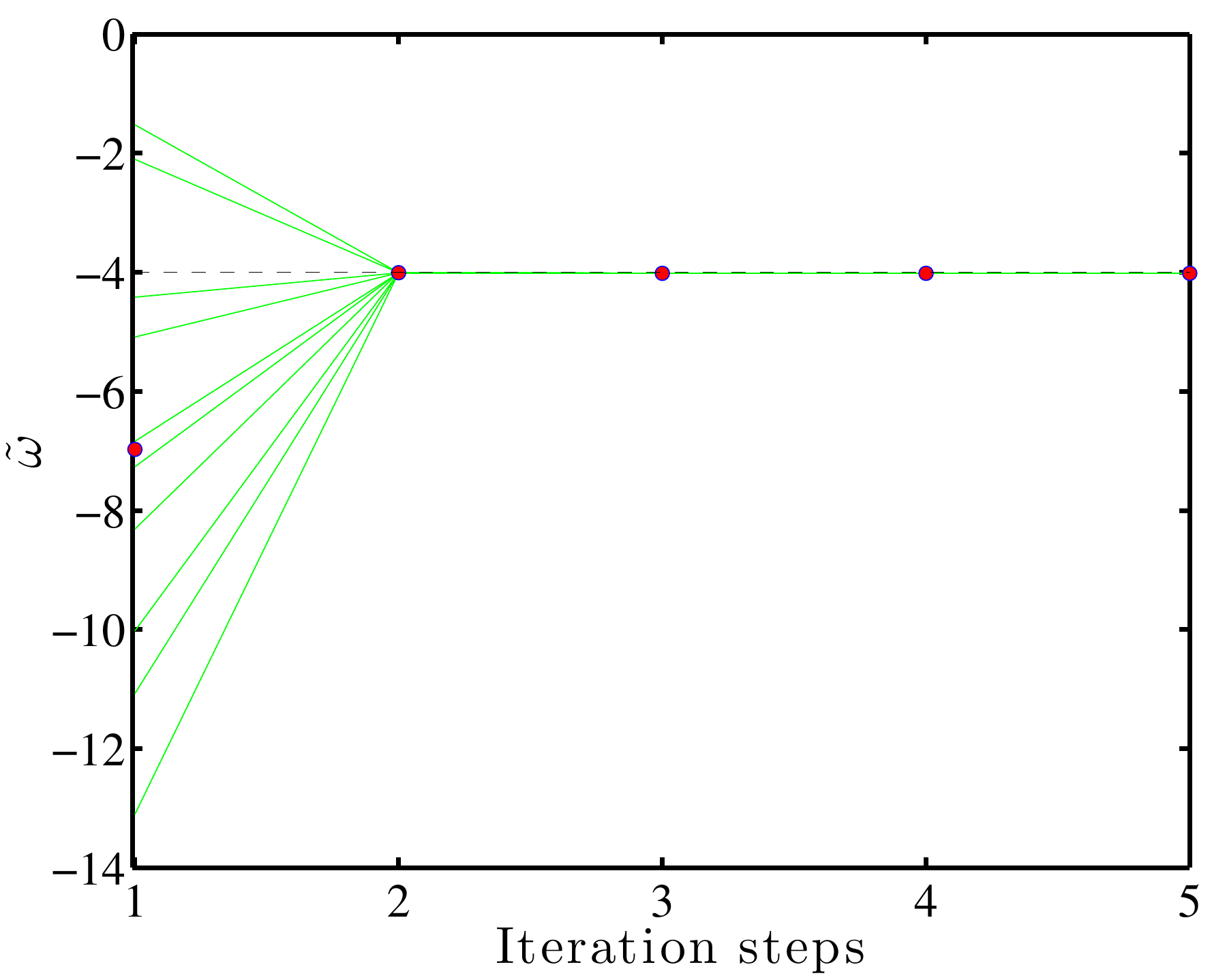}}\\
	\end{center}
	\caption{ Convergence history of the inferred parameter $\tilde{\omega}$ of the drag force 
		distribution in the verification case. Synthetic truth of $\tilde{\omega}$ is $-4$, to which the 
		inverse modeling scheme is blind. The prior of $\tilde{\omega}$ is unbiased
		($\mathop{\mathbb{E}}(\tilde{\omega}) = -4$) in panel (a), while it is biased 
		($\mathop{\mathbb{E}}(\tilde{\omega}) = -7$) in panel (b), where 
		$\mathop{\mathbb{E}}(\cdot)$ denotes expectation. Note that all the samples converge
		and overlap with each other after the first iteration.
	}
	\label{fig:para_syn}
\end{figure}

The convergence histories of parameter $\tilde{\omega}$ in the two scenarios, 
starting from the unbiased prior and biased prior, are shown in 
Figs.~\ref{fig:para_syn}a and~\ref{fig:para_syn}b, respectively. 
The synthetic truth $\tilde{\omega} = -4$ is also plotted for comparison.
The samples of $\tilde{\omega}$ initially scattered across
the range from -8 to 0 converge to the synthetic truth quickly (Fig.~\ref{fig:para_syn}). 
Each sample of $\tilde{\omega}$ is corrected based on
the comparison of simulated and observed velocities and TKE. 
For the case with unbiased prior, the performance of the proposed
method is excellent. However, in practice the truth is not known a priori,
and thus the prior distribution is often biased. The second case
is representative of the latter, more realistic scenario. 
For the initial samples with a biased mean and larger interval, the convergence
to the truth is also achieved rapidly. All the samples and their mean are 
corrected to $\tilde{\omega} = -4$. It is worth noting that the convergences 
are achieved within only one iteration for both scenarios, which is 
because the dimension of parameter space is relatively low, and the 
observation data are sufficient. The importance of the prior diminishes 
increasing observation data. The agreement between the inferred result and 
the synthetic truth demonstrates the merits of the proposed inverse modeling 
scheme.   

\subsection{Case with Experimental Data}
\label{sec:result2}  
A realistic inverse modeling case with the real experimental data 
(introduced in Sec.~\ref{sec:prob}) is explored. As mentioned above, the body 
force $f$ caused by the porous disk is represented by a number of basis functions
with corresponding coefficients (see Eq.~\ref{eq:para} ), which are to be
inferred to obtain an improved flow field prediction. In this study $M = 3$ 
basis functions are used to represent the force distribution.
We specified a uniform distribution within $[-10, 10]$ for the prior of each
coefficient to be inferred. The magnitude of drag force is $4.5 \times 10^{-3}$ N, 
which is measured from the experiment. The aim here is to infer how this drag
force is distributed over the disk. We use the same sample size of $N = 10$ 
as that used in the synthetic case. 
\begin{figure}[!h]
	\begin{center} 
		\subfloat[Convergence history of $\omega_1$]{\includegraphics[width=0.5\textwidth]{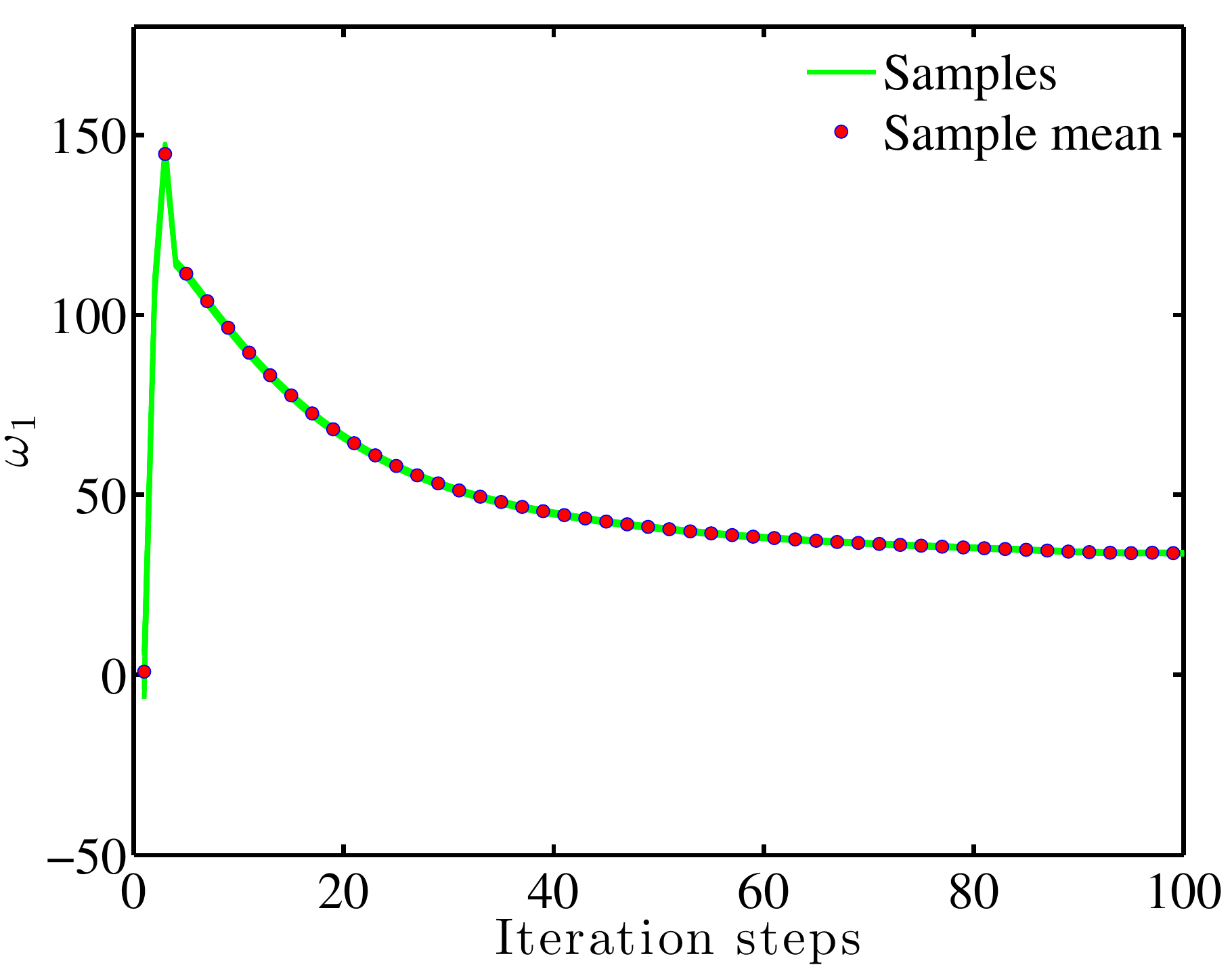}}
		\subfloat[Convergence history of $\omega_2$]{\includegraphics[width=0.5\textwidth]{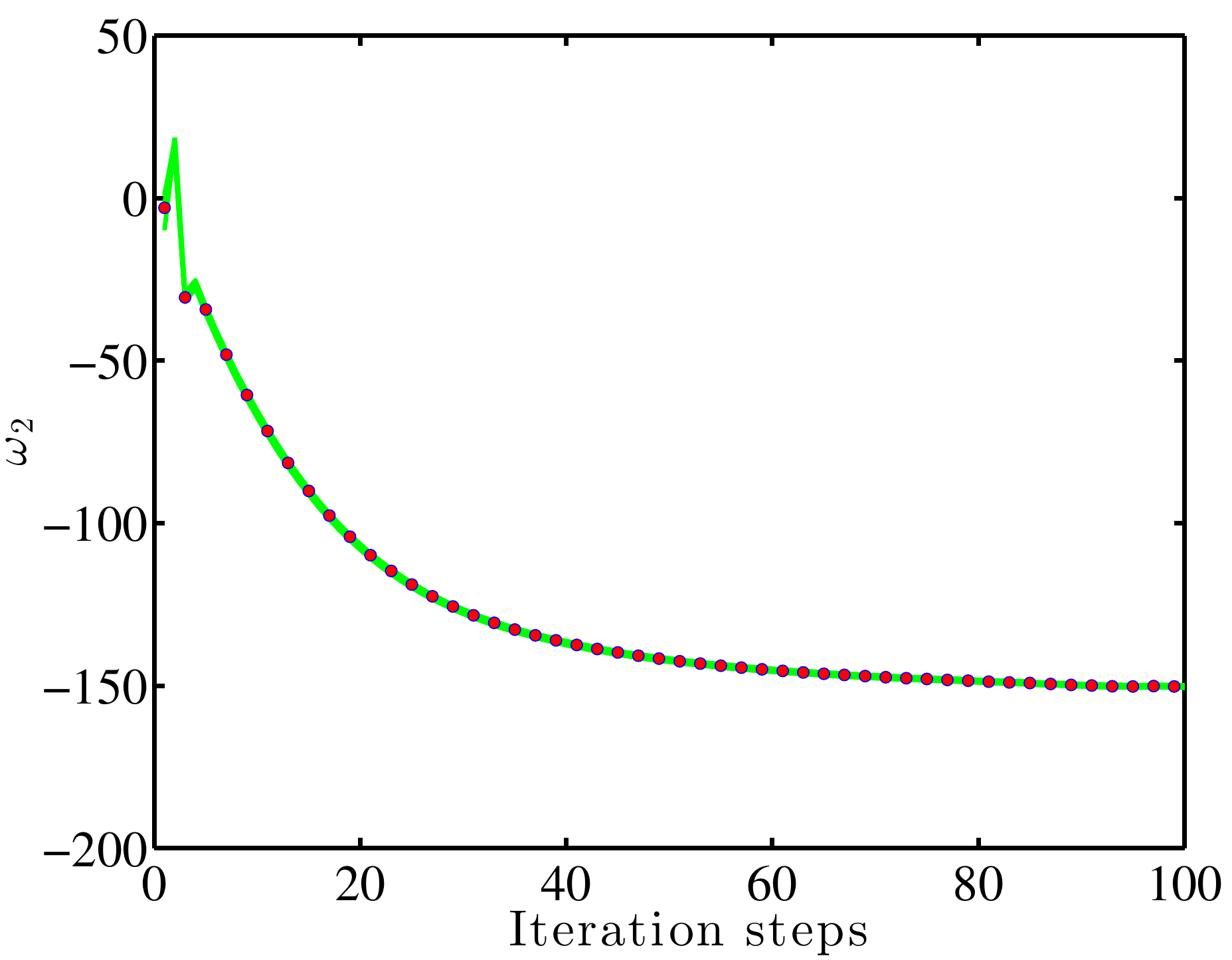}}\\
		\subfloat[Convergence history of $\omega_3$]{\includegraphics[width=0.5\textwidth]{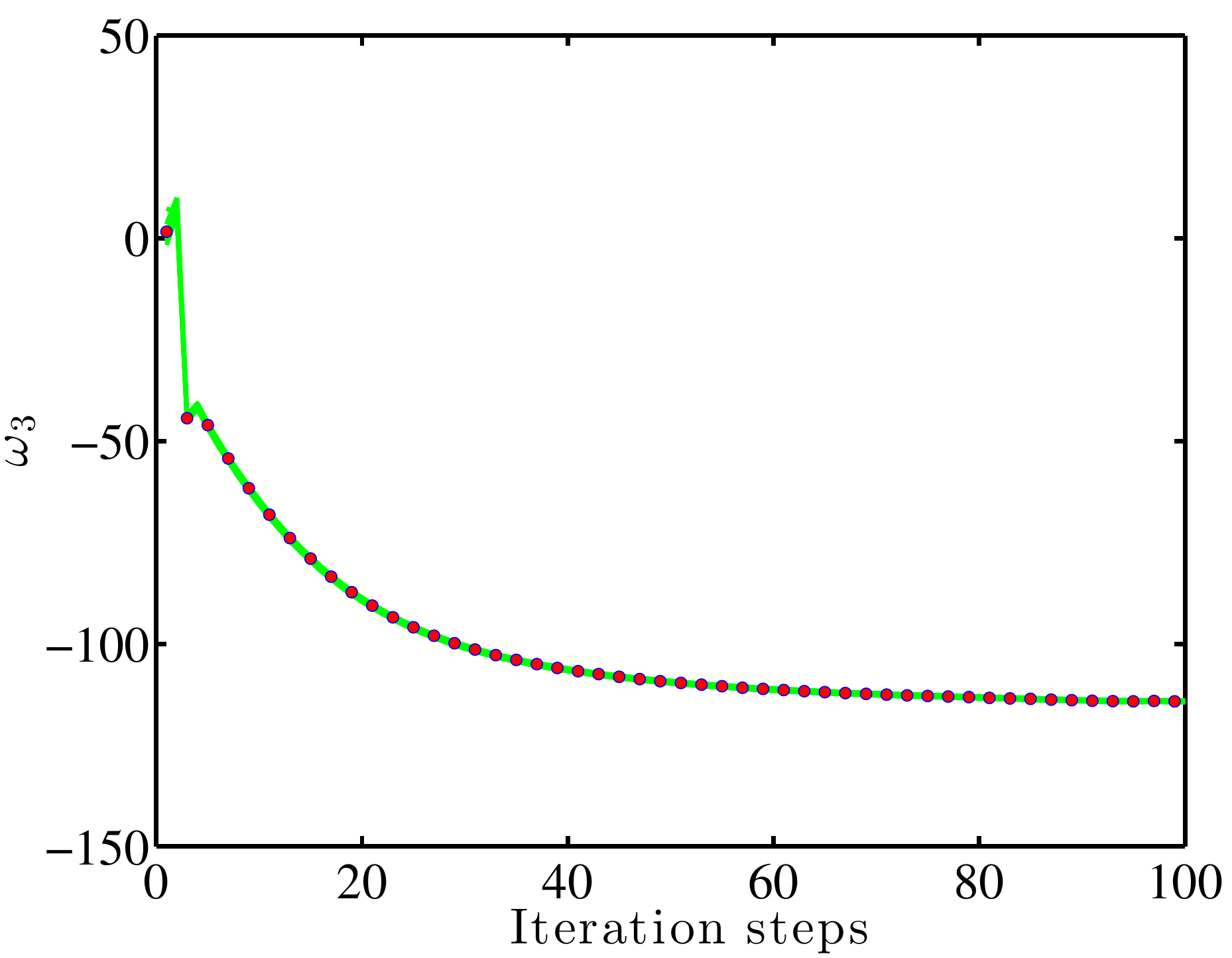}}
	\end{center}
	\caption{
		Convergence histories of the inferred parameters $\omega_1$, $\omega_2$ 
		and $\omega_3$ for the drag force distribution with experimental data. 
		Note that the samples converge and overlap with each other after the first
		iteration.}
	\label{fig:para_real}
\end{figure} 

The convergence histories of coefficients $\omega_1$, $\omega_2$, and
$\omega_3$ are shown in Figs.~\ref{fig:para_real}a,~\ref{fig:para_real}b, 
and~\ref{fig:para_real}c, respectively. Similar to the synthetic case, the samples 
converge quickly for all the coefficients. However, a notable difference 
is that sample means still change with the iterations. After about 50 iterations,
all the coefficients converge. Compared to the synthetic case, the convergences 
based on the real experimental data need more iterations. Because of 
the various approximations in the numerical modeling (e.g., turbulence modeling) and 
the measurement errors in the experimental data, a force distribution that makes 
the model predictions to exactly agree with the experimental observations may not exist.  
This is in stark contrast to the synthetic data, where the data are obtained 
by assuming a force distribution. Therefore, more iterations are needed to achieve 
statistical convergence of the samples.

\begin{figure}[htbp]
	\begin{center}
		\includegraphics[width=0.5\textwidth]{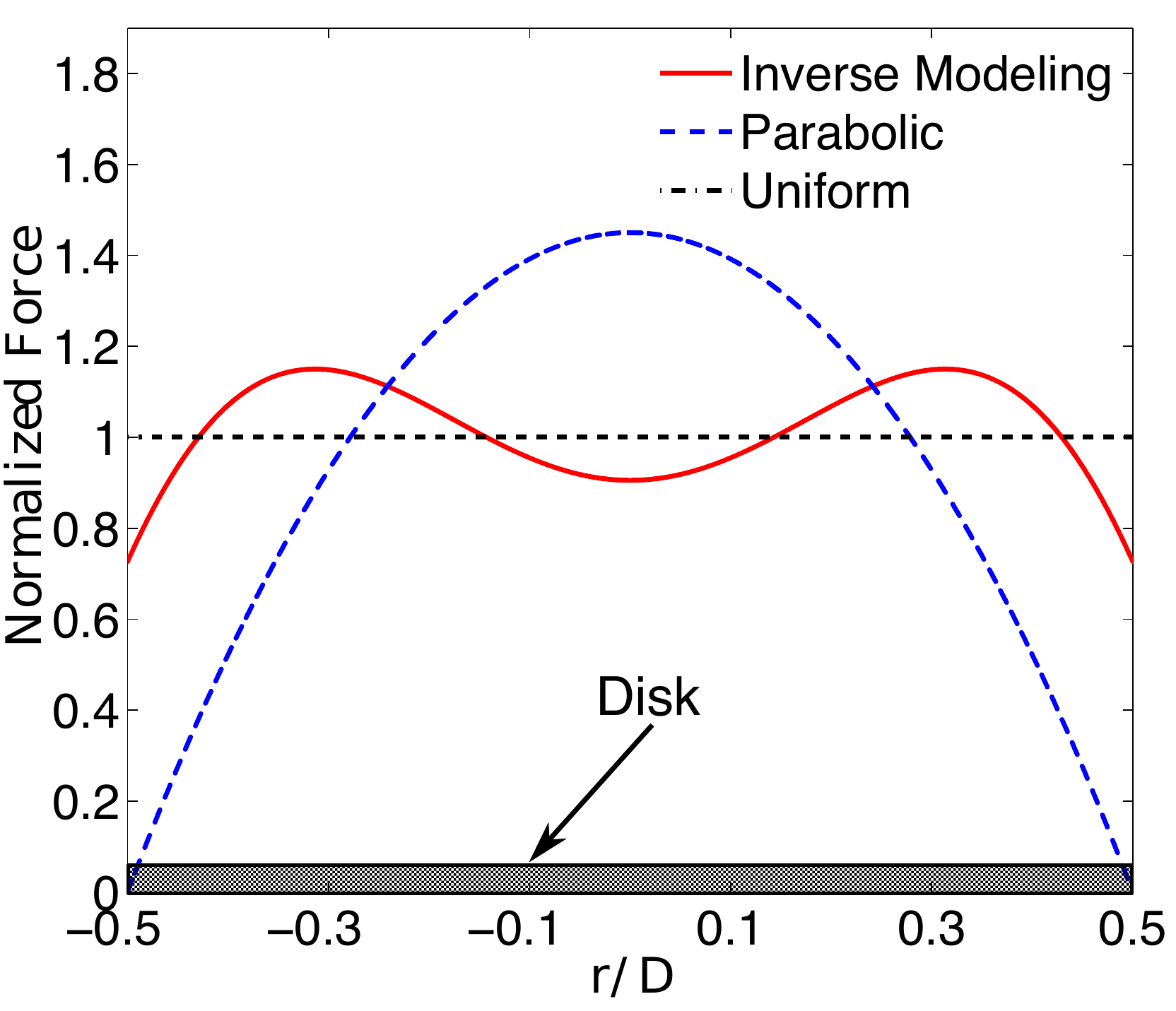}
		\caption{Estimated force distribution with the inferred parameters. 
					  The parabolic force distribution and the uniform force distribution 
					  are plotted for comparison.
		}
		\label{fig:disOth1}
	\end{center}
\end{figure}

The force distribution reconstructed with the converged coefficients 
(sample means after 100 iterations) is shown in Fig.~\ref{fig:disOth1}. 
Two commonly assumed force distributions in literature are also
plotted in the same figure for comparison. One is that the drag force 
is assumed to be uniformly distributed over the disk, which is frequently
used in the literature~\cite{jimenez2007advances, jimenez2008large, 
	wu2011large}. The other is to assume that porosity of the disk is zero, and
thus the force distribution is parabolic based on the potential flow 
theory~\cite{currie2012fundamental}. The sideview of the disk are plotted 
with a dashed rectangular along the horizontal axis to facilitate interpretation. 
Based on the inversion results, the force is non-uniformly distributed on 
the disk with normalized magnitude from 0.7 to 1.2 (red/thick line), spatially varying 
around the uniformly distributed force (black/dot-dashed line). The force magnitude
is smallest at the edge of the disk ($r/D = 0.5$), while it peaks at near the 
half of its radius ($r/D = 0.3$). Towards center of the disk ($r/D = 0$), 
the force decreases and reaches a valley. Compared
to the parabolic force distribution (blue/dashed line), the estimated force does 
not have a significant decrease at the edge of disk. Another notable difference 
lies on the peak of the force, which is not at the disk center for the inversion 
result. Due to the porous structures, the force distribution of the porous disk is 
different from that of the solid one. 

\begin{figure}[!h]
	\begin{center} 
		\subfloat[Along the disk radius at $x/D = 2.11$]{\includegraphics[width=0.5\textwidth]{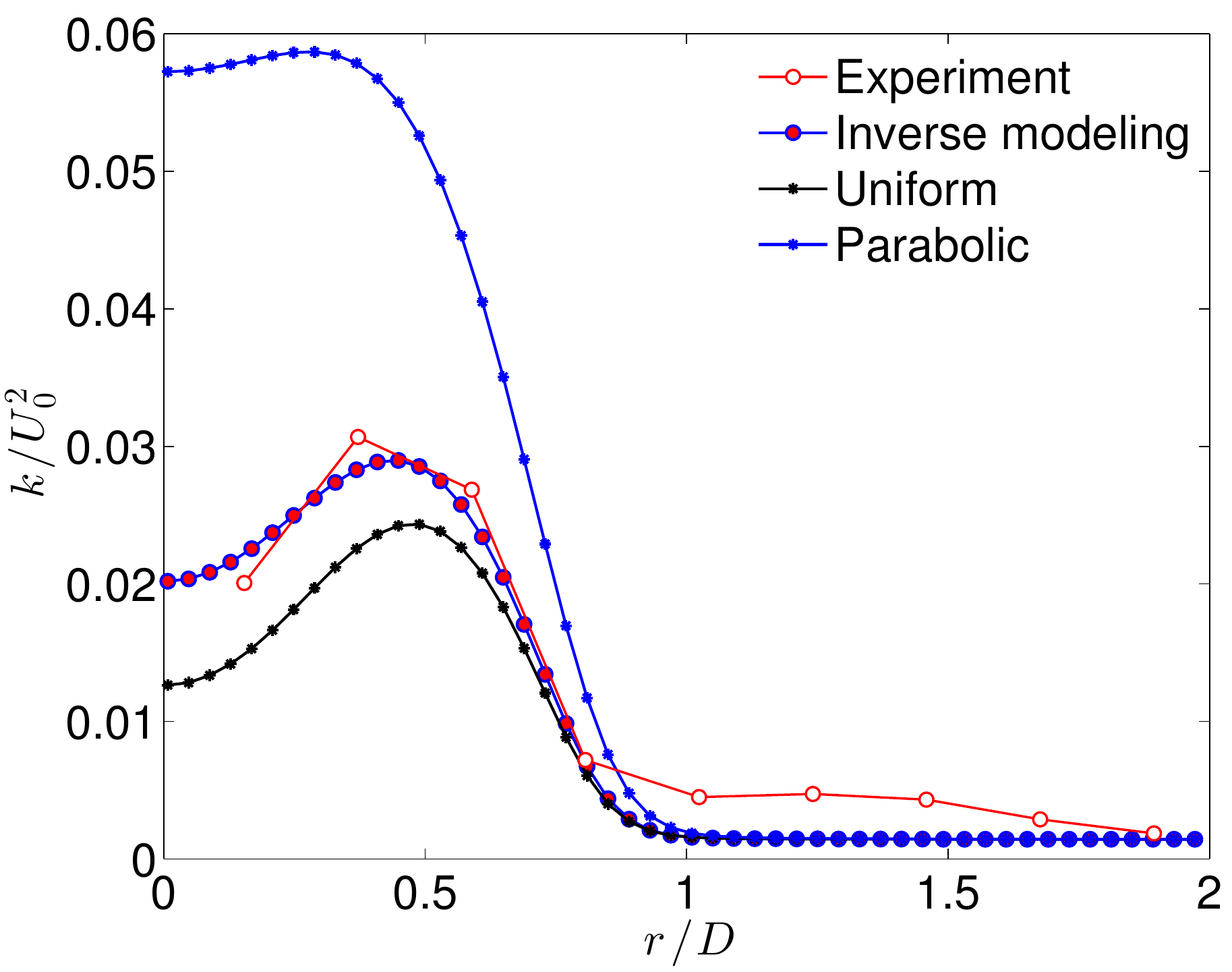}} 
		\subfloat[Along the disk radius at $x/D = 2.98$]{\includegraphics[width=0.5\textwidth]{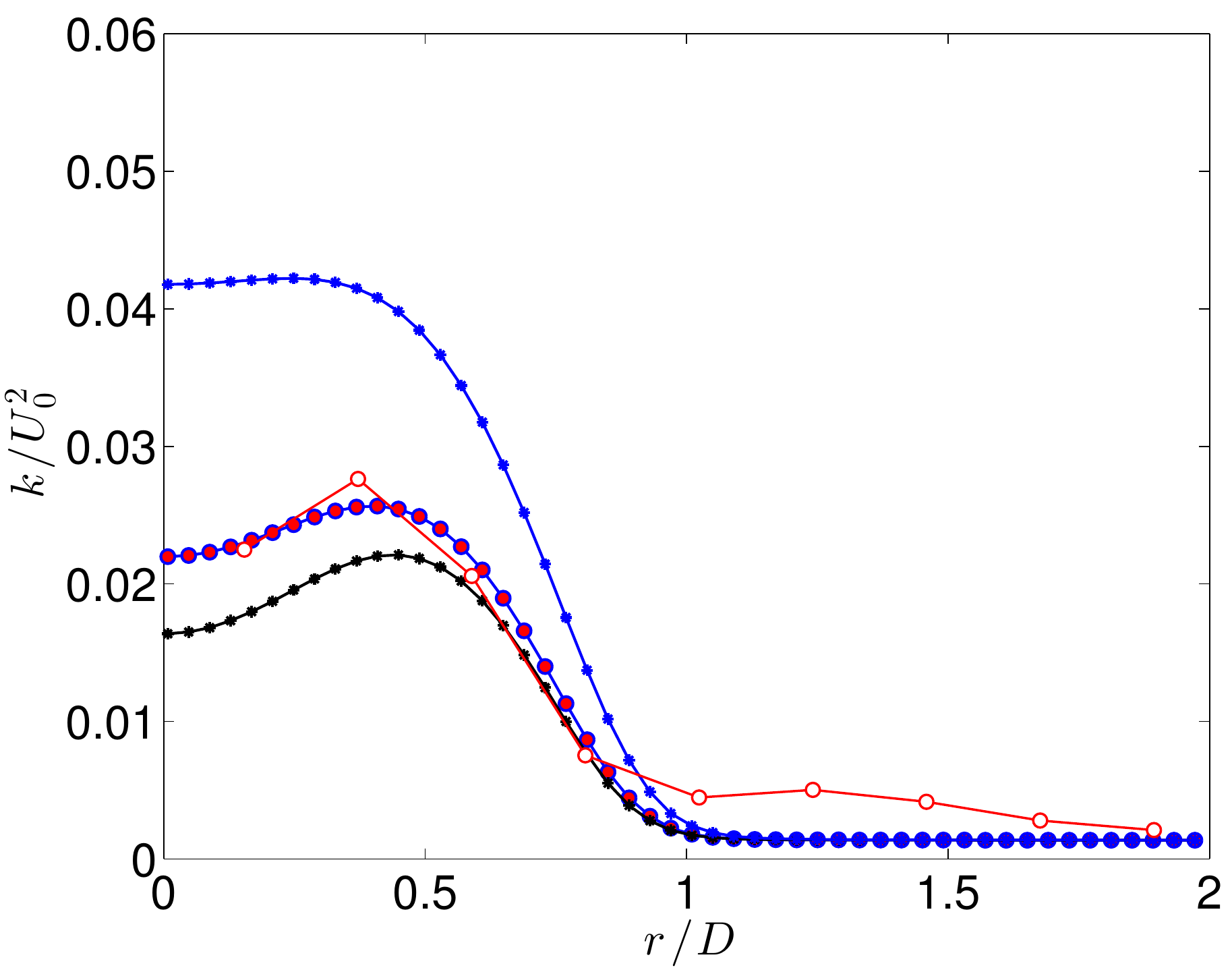}}\\
		\subfloat[Along the streamwise at $r/D = 0.154$]{\includegraphics[width=0.5\textwidth]{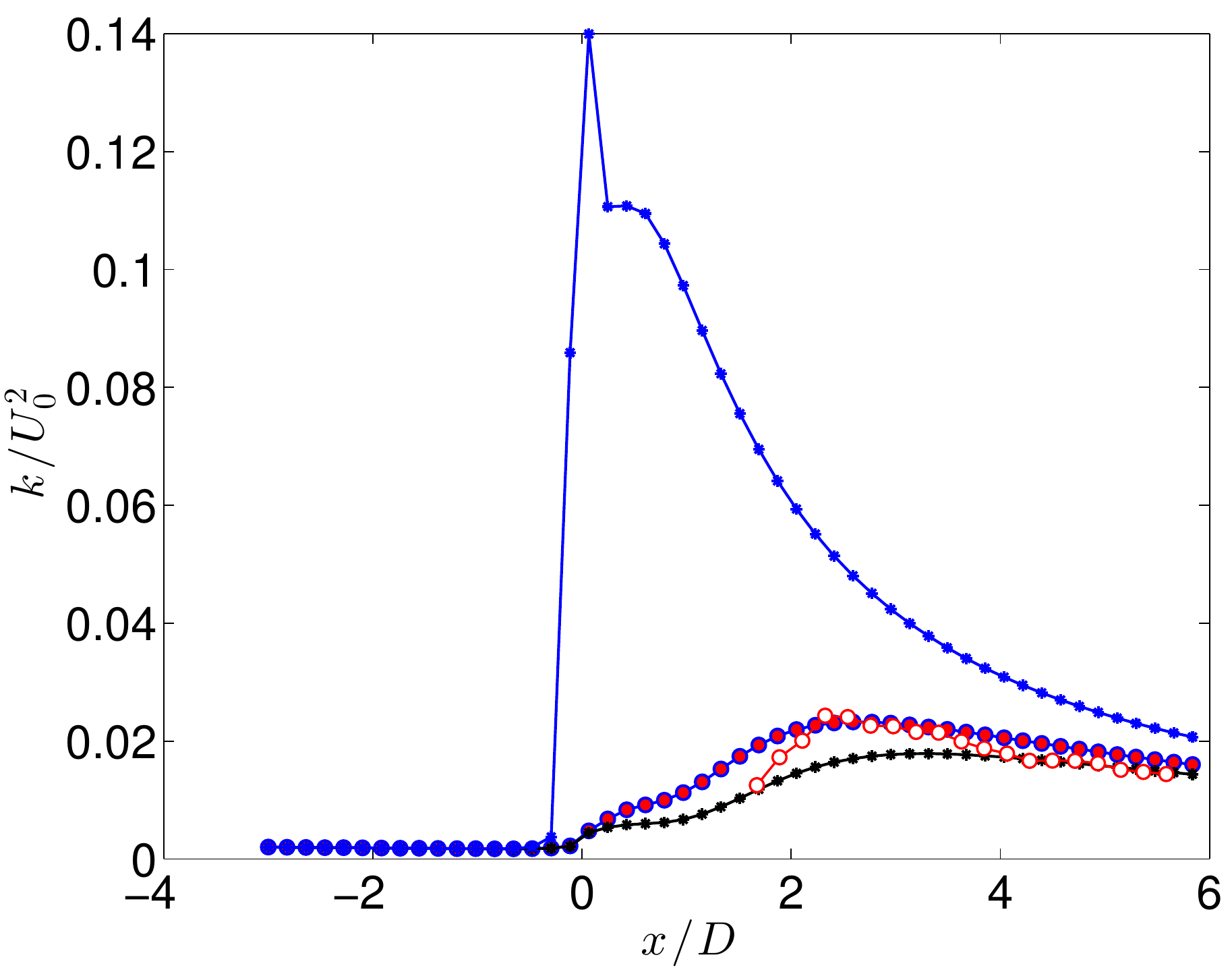}} 
		\subfloat[Along the streamwise at $r/D = 0.372$]{\includegraphics[width=0.5\textwidth]{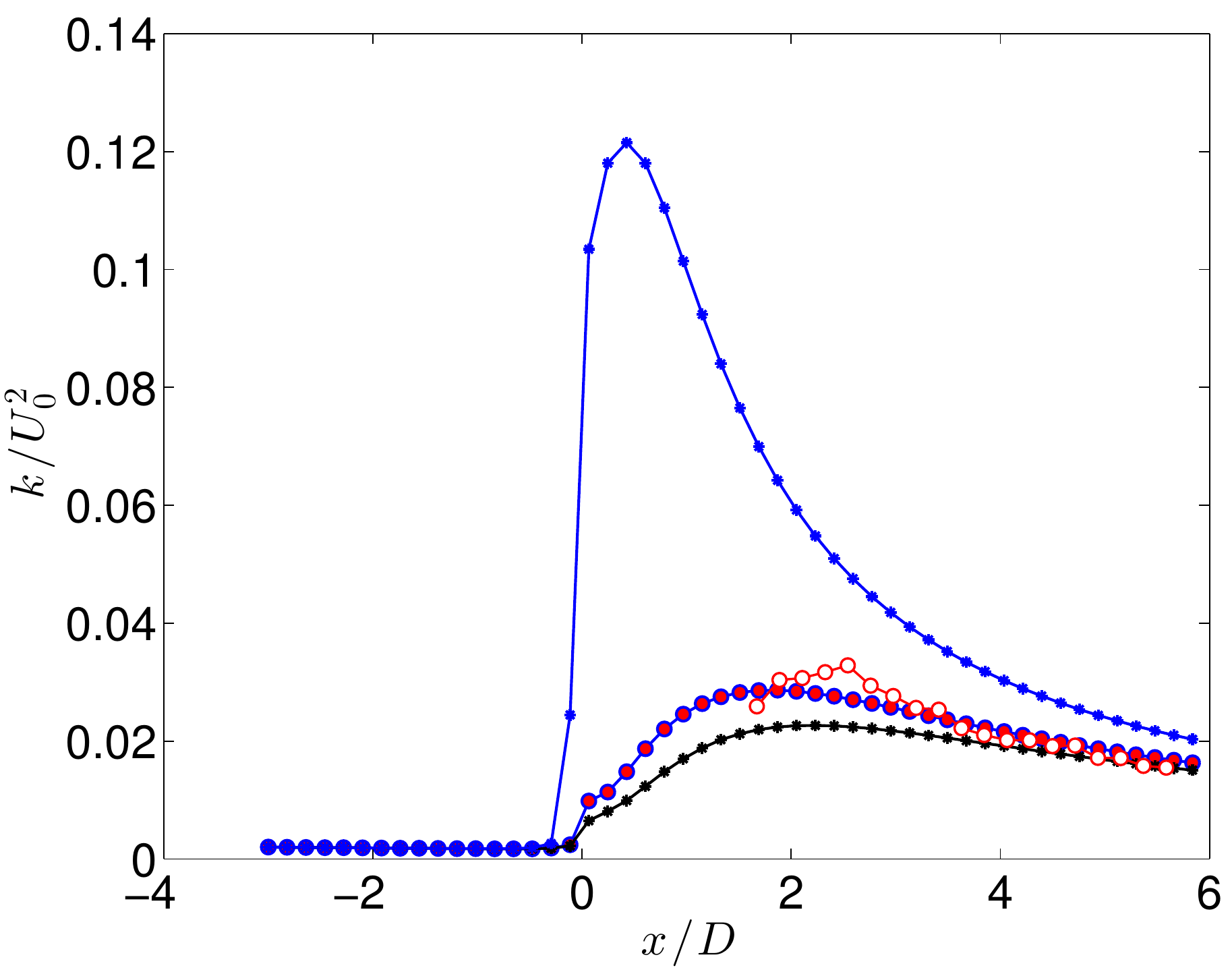}}
	\end{center}
	\caption{Comparison of turbulence kinetic energy (TKE) profiles obtained from the
		experiment and numerical simulations along the radial lines at (a) $x/D = 2.11$, 
		(b) $x/D = 2.98$ and along the streamwise lines at (c) $r/D = 1.54$, (d) $r/D = 3.72$. 
	}
	\label{fig:realflowk1}
\end{figure} 

As the inference is based on the experimental data of the flow field, 
there is no ground truth to directly validate the inferred force distribution. 
Therefore, we have to evaluate it with corresponding model predictions. 
That is, the inferred force distribution is applied to perform a forward simulation, 
and the simulated flow quantities (e.g., the velocity and TKE) are compared
with the experimental data to evaluate the inferred force distribution.  
The comparisons of the TKE profiles with the experimental data along the 
spanwise and streamwise directions are shown in Fig.~\ref{fig:realflowk1}. 
The simulated results with uniformly and parabolically distributed forces are 
also plotted for comparison. We only shown the results of two radial lines 
at $x/D = 2.11$ and $x/D = 2.98$ (Figs.~\ref{fig:realflowk1}a and~\ref{fig:realflowk1}b),
and two streamwise lines at $r/D = 0.154$ and $r/D = 0.372$ 
(Figs.~\ref{fig:realflowk1}c and~\ref{fig:realflowk1}d). The results at other 
locations have similar characteristics and thus are omitted for brevity. 
We can see that the TKE profiles obtained with the inferred force distribution 
has a better agreement with the experimental data compared
to the results with the other two force distributions, especially in the wake behind the 
disk ($r/D = 0 \sim 0.5$). In contrast, the TKE is overestimated when the 
parabolic force distribution is applied, while it is underestimated as the uniform 
force distribution is used. In the wake at $2.11D$ after the disk, the TKE obtained 
with parabolically distributed force is about twice larger than the inversion result and 
the experimental data (Fig.~\ref{fig:realflowk1}a), while the overestimation is reduced
at $2.98D$ after the disk (Fig.~\ref{fig:realflowk1}b). On the contrary, the TKE 
obtained with the uniformly distributed force is slightly smaller than the inversion 
result and experimental data. These trends can be also clearly seen from
the TKE profiles along the streamwise directions (Fig.~\ref{fig:realflowk1}d 
and Fig.~\ref{fig:realflowk1}d). The results shown above indicate that the 
force distribution largely affects the simulated TKE, and the differences of TKE 
simulated with different force distributions are reduced in the far wake far region. 
These differences can be explained based on the momentum equation
(Eq.~\ref{eq:common-u}). For the parabolic force distribution, the 
gradient~$\frac{\partial f_i}{\partial x_j}$ of the body force is overestimated 
because the porous structure is ignored. Overestimated force gradient 
amplifies the velocity gradient, which in turn leads to an increased production term 
in the TKE equation (Eq.~\ref{eq:k}), i.e., $\tau_{ij}\partial U_i/\partial x_j$.
Therefore, the turbulence kinetic energy is overestimated. 
On the contrary, uniform force distribution underestimates 
turbulence kinetic energy, since the force gradient is assumed to be zero. 
In contrast, the inferred force distribution markedly improves the simulation results,
though discrepancies still can be seen in near-wall region ($r/D > 1.0$), which
might be because of the influence of wall effect of the channel in the experiment. 
The comparisons results shown above demonstrate merits of the proposed 
data-driven inverse modeling.   

\begin{figure}[!h]
	\begin{center} 
		\subfloat[Along the radial direction at $x/D = 2.11$]
		{\includegraphics[width=0.5\textwidth]{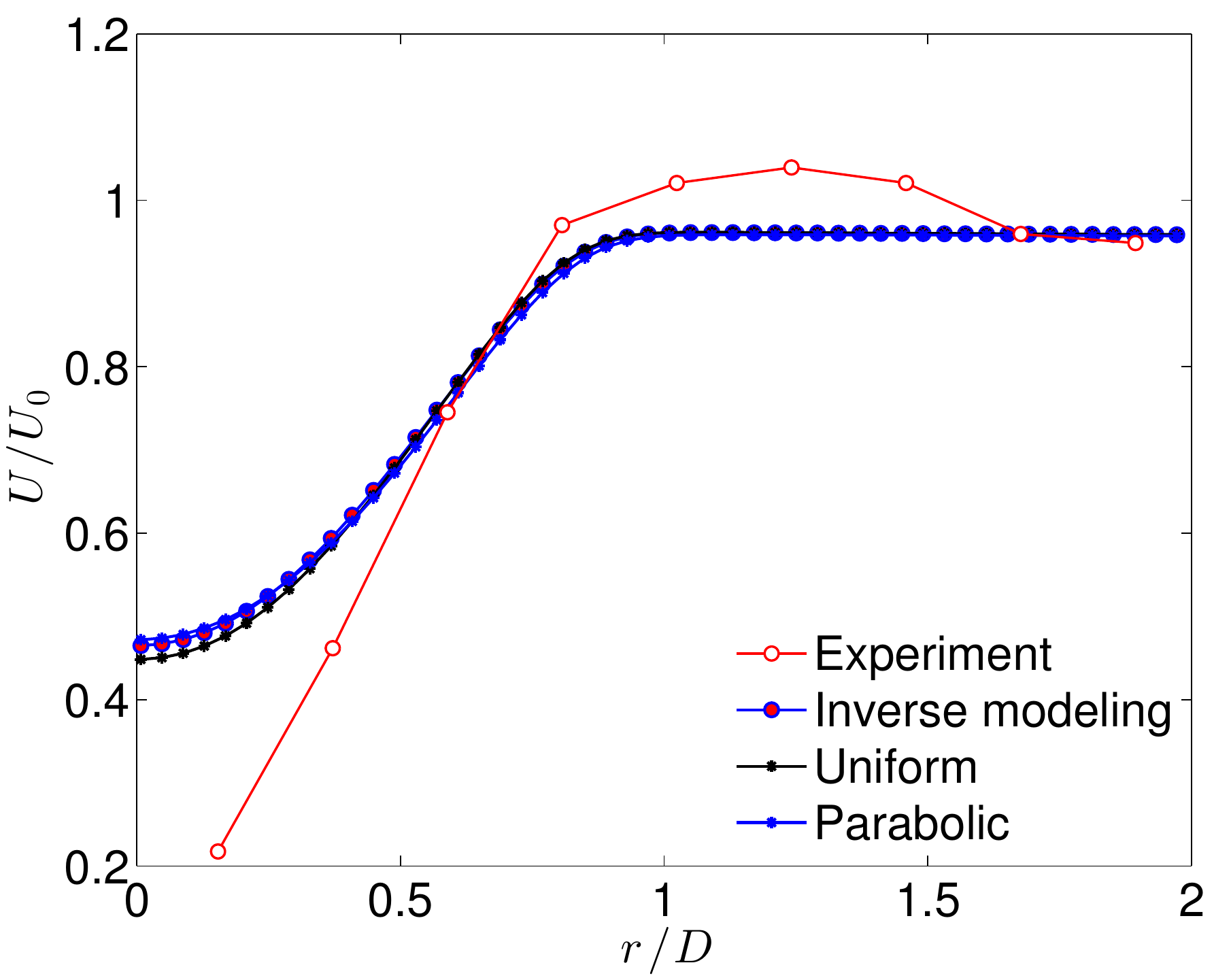}} 
		\subfloat[Along the disk radial direction at $x/D = 2.98$]
		{\includegraphics[width=0.5\textwidth]{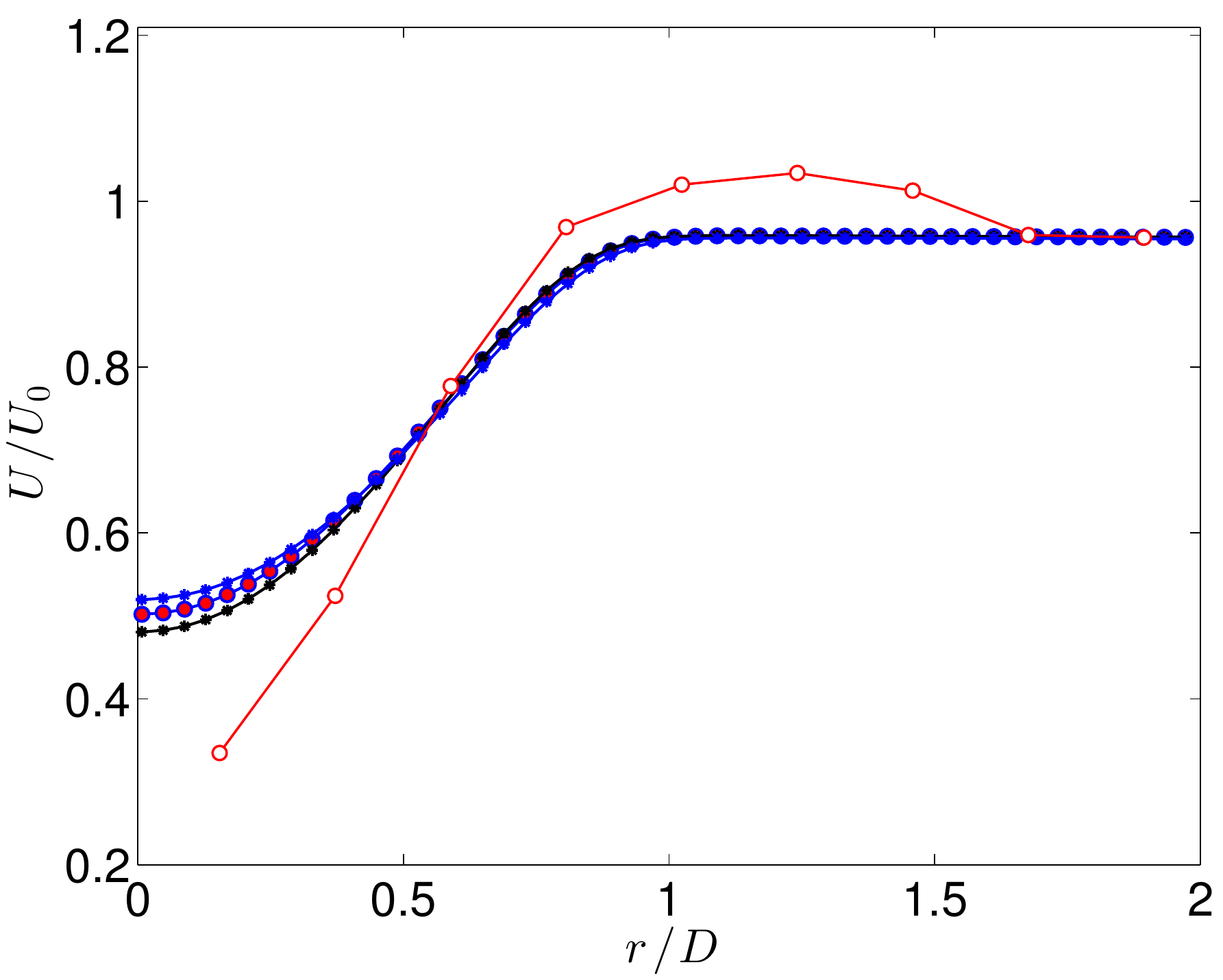}}
	\end{center}
	\caption{Comparison of the mean velocity profiles obtained with different force distributions (uniform, parabolic, and inferred) along the disk radial direction at (a) $x/D = 2.11$ and (b) $x/D = 2.98$. The experimental data are also plotted for comparison. Only the force distribution is calibrated in the inversion case.
	}
	\label{fig:realflowU1}
\end{figure}

For the velocity prediction, a large velocity deficit exists in the wake of the disk, and it
is reduced in the far wake. However, the simulated results under-predict the velocity deficit 
compared to the experimental data (Fig.~\ref{fig:realflowU1}).
The mean velocity simulated with the inferred force distribution does not shown noticeable
improvement, and the velocity profiles obtained with differently distributed
forces overlap with each other, indicating that the force distribution has less influence on the
velocity field of the wake. Unlike the synthetic case, more uncertain factors existed
in the realistic scenarios other than the force distribution may cause this discrepancy.
For example, the measured drag force $f_t$ has uncertainties; the chosen coefficients $C_1$ 
and $C_2$ in the $k$--$\varepsilon$ turbulence model may not be optimal for this particular flow.
In current proposed inverse modeling scheme,
all these uncertain factors also can be considered as parameters and augmented into the state, 
and thus they can be corrected along with the other state variables based on observation data. 
We demonstrate this by the following case, where the force magnitude $f_t$
and the $k$--$\varepsilon$ model constants are also considered as uncertain parameters 
to be inferred as well as the force distribution. For the $k$--$\varepsilon$ model constants, 
we find $C_1$ is most sensitive to the flow prediction for this particular flow. 
Therefore, for the turbulence model parameters, only $C_1$ is to be inferred. 
The priors of the force magnitude $f_t$ and turbulence
constant $C_1$ are perturbed based on $\bar{f}_t = 0.45$ N and $\bar{C}_1 = 1.44$, 
which are the mean value of measurement data of $f_t$ and the standard value for
 $C_1$~\cite{wilcox98}, respectively. The initial samples of $f_t$ and $C_1$ are 
uniformly drawn from $0.45 \pm 0.015$N and $1.44 \pm 0.2$, respectively.
The priors of force distribution parameters and other computational setup are as the same as those
of the case shown above.  After 100 iterations, the ensemble converges statistically. The inferred
force distribution is similar as the one shown in Fig.~\ref{fig:disOth1}. The
inferred force magnitude $f_t$ increases to $4.57 \times 10^{-3}$ N, while the inferred 
turbulence constant $C_1$ is $1.56$, which is slightly larger than the standard value. 

\begin{figure}[!h]
	\begin{center} 
		\subfloat[Along the radial direction at $x/D = 2.11$]
		{\includegraphics[width=0.5\textwidth]{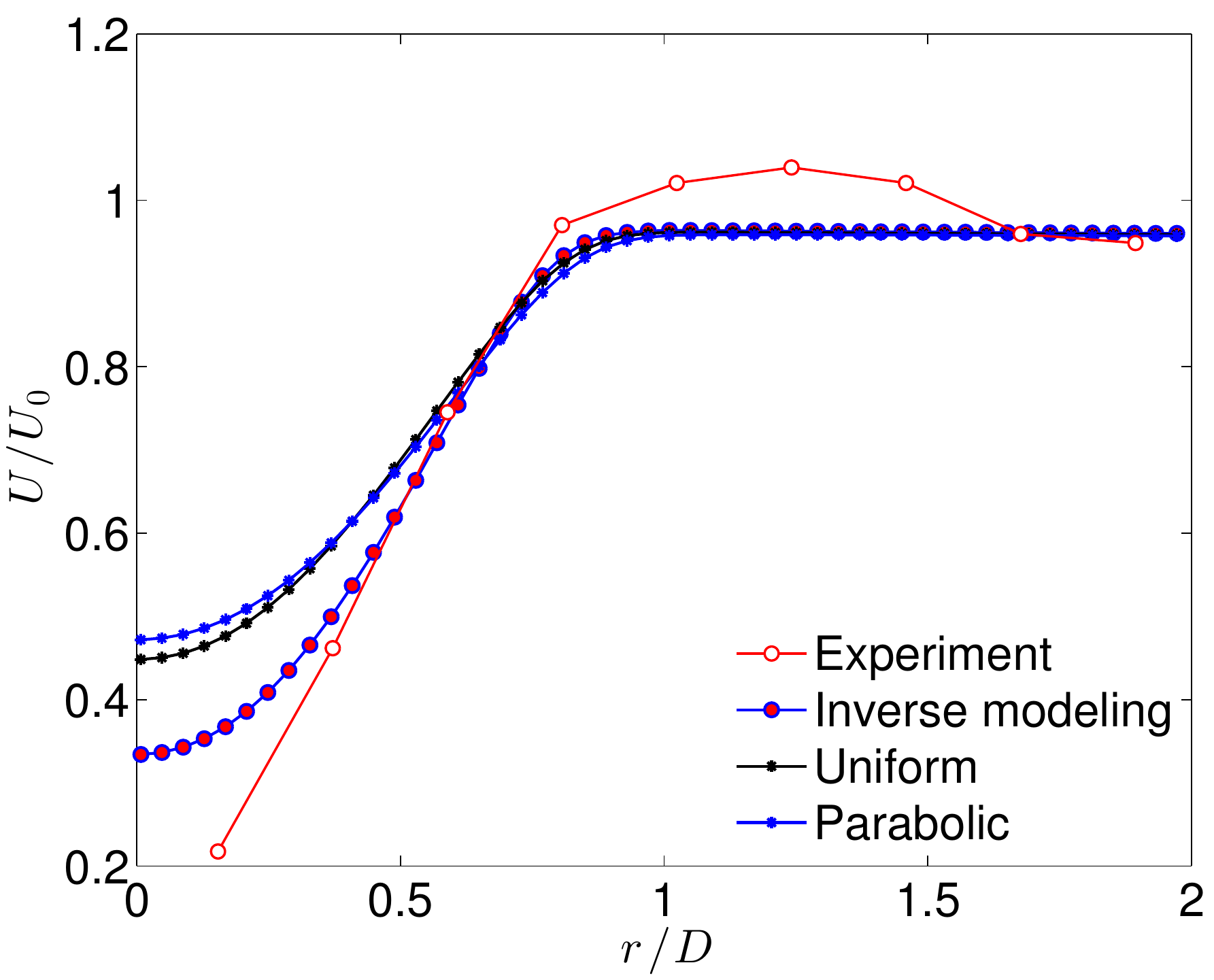}} 
		\subfloat[Along the radial direction at $x/D = 2.98$]
		{\includegraphics[width=0.5\textwidth]{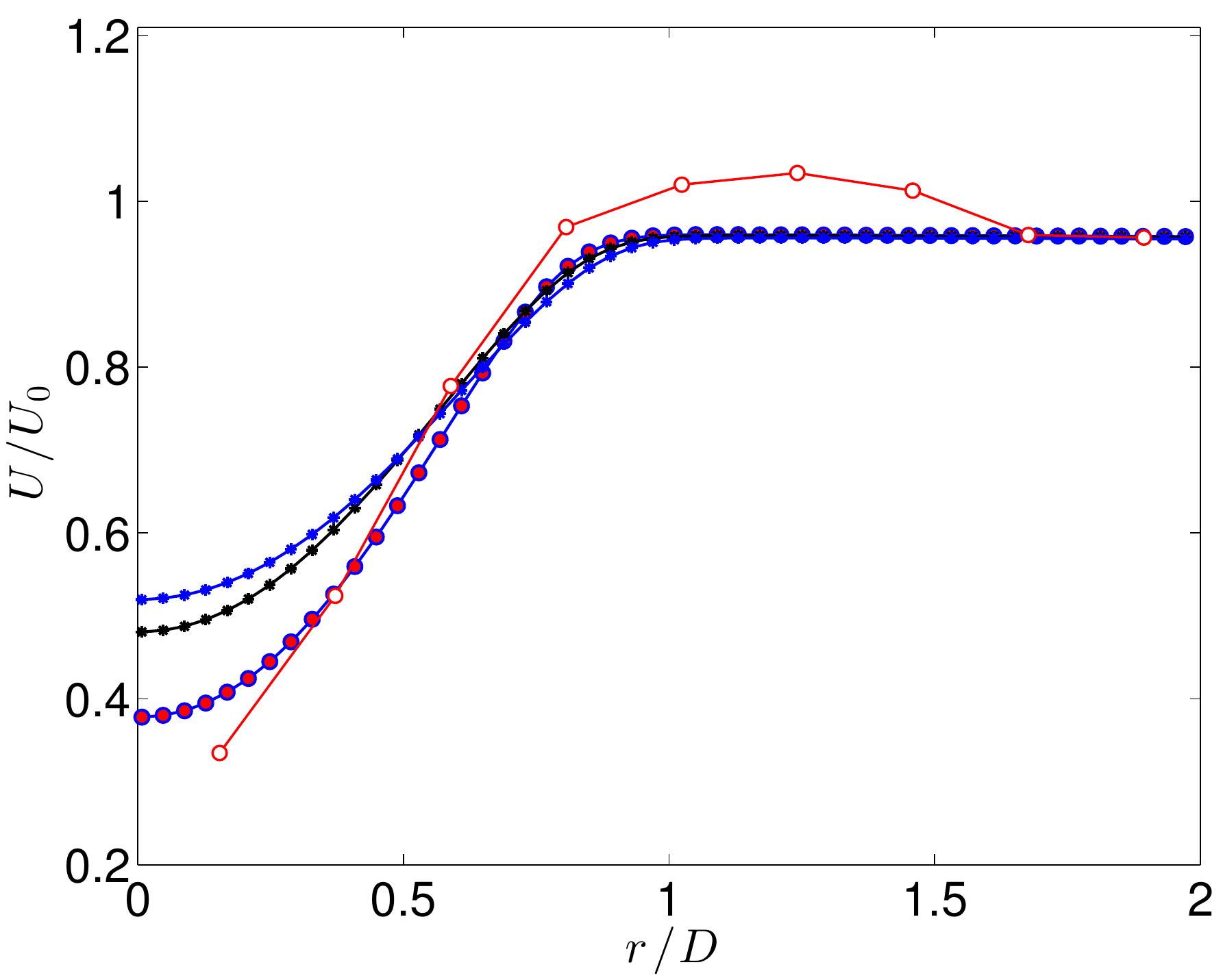}}\\
	\end{center}
	\caption{Comparison of the mean velocity profiles obtained with different 
		force distributions (uniform, parabolic, and inferred) along the 
		radial direction at (a) $x/D = 2.11$ and (b) $x/D = 2.98$. The experimental 
		data are also plotted for comparison. The force distribution, 
		force amplitude, and turbulence model are calibrated in the inversion case.}
	\label{fig:realflowU2}
\end{figure}

With consideration of the uncertainties in the force distribution, force magnitude and turbulence
model, both the predicted turbulence kinetic energy and velocity are significantly improved. For the
turbulence kinetic energy, the predictions from inverse modeling well agrees with the experimental
data, which is similar as those shown in Fig.~\ref{fig:realflowk1} and thus is omitted for
brevity. Meanwhile, the velocity predictions are also markedly improved (Fig.~\ref{fig:realflowU2})
compared to the inversion results by only considering the uncertainties in force distribution
(Fig.~\ref{fig:realflowU1}). The velocity recovery in the wake, which is over-predicted in
Fig.~\ref{fig:realflowU1}, has been captured more accurately here (Fig.~\ref{fig:realflowU2}).
{\color{black} It was observed in the literature that RANS solvers based on standard turbulence models
  tend to overpredict the wake recovery when used in conjunction with the actuation disk model
  (i.e., representing disks/turbines as momentum sinks). This deficiency was attributed to the fact
  that standard RANS models were not able to account for the energy transfer between the large scale
  turbulence associated with the coherent structures to small scale turbulence in the dissipation
  range, and the energy transfer mechanism is dominant in the turbine wake
  problem~\cite{shives2016adapted,el2008extended}.  Variants of the standard $k$--$\varepsilon$
  model have been proposed in the turbine wake modeling community to overcome this
  difficulty~\cite{shives2016adapted}.  For example, in the $k$--$\varepsilon$--$S_\varepsilon$
  model proposed by El Kasami and Masson~\cite{el2008extended}, a source term $S_\varepsilon$ is
  added to the transport equation for the dissipation rate $\varepsilon$ to obtain wake recovery
  rates in better agreement with utility scale turbine data. As mentioned earlier, $C_1$ in the
  $\varepsilon$ equation (\ref{eq:eps}) is associated with the production term of the dissipation
  rate. Therefore, using a larger value of $C_1$ as inferred from the wake velocity data in this
  work has a similar effect of boosting the production as adding a source term in the $\varepsilon$
  equation. In addition, accounting for the uncertainties in the measured force may also have
  contributed to the improved results as shown above, but its effects on the wake recover rate is
  likely to be minor.  By accounting for these potential uncertainties in the model and in the
  experimental data and by properly representing and correcting them, the simulation results are
  markedly improved and better agree with the experimental data. We emphasize that the number of
  dimensions of the uncertainty space is large, particularly for the uncertainty in the drag force
  distribution on the disk. Consequently, it is not feasible to tune the coefficients manually and
  empirically to achieve better agreement with the experimental data.  The results above
  demonstrates the performance of the proposed inverse modeling scheme in realistic applications
  with experimental data. }



\section{Discussion}
\label{sec:dis}
In this section, we further discuss the merits and limitations of the proposed method.

\subsection{Significance and Potential of the Data-Driven Approach}

{\color{black} With the data-driven inverse modeling approach, the velocity and turbulent kinetic
  energy measurement data in the wake of a disk are used to infer the drag force incurred by the
  disk, or more precisely, the drag force used to represent the hydrodynamic effects of the
  disk. This non-traditional modeling approach and its significance deserve further
  clarification.  We consider two scenarios as discussed in Section~\ref{sec:intro}. First, when the
  data-driven inverse modeling approach as presented above is utilized to provide forecasting of a
  farm of wind or tidal turbines, usually only sparse measure data (e.g., velocity and turbulence
  levels) are available. These measurements are used to infer the force distribution, which further
  provides the full field velocity, turbulence, and other quantities of interest. Second, when the
  data driven approach is used to calibrate ad hoc model or guide the developments thereof,
  measurement data can be available in a small number of cases, and predictions are sought for other
  cases without data (e.g., those with different turbines or at different operation conditions from
  the cases where measurements were taken). In this scenario, the spatial distribution of the drag
  can be extrapolated from the calibration cases to the prediction cases, where an implicit
  assumption is that the flow pattern between the calibration and predictions cases are indeed
  similar. This assumption are often valid. The merits of the present method have been demonstrated
  in both scenarios in the context of reducing model-form uncertainty in RANS
  models~\cite{xiao-mfu,wu2015bayesian}, where the Reynolds stress fields and the full field
  velocity are inferred from sparse measurements by using the inverse modeling method present in
  this work. Finally, if the bimodal spatial distribution of drag forces as shown in
  Fig.~\ref{fig:disOth1} is universal among a number of turbines and flow conditions, the
  traditional actuation disk models can be adapted to yield such drag distributions.  An important
  novelty of the present inverse modeling approach is that it leads to results that lend clear
  physical interpretations to model developers in application domains. These results are more likely
  to be universal than those obtained by using the parametric approach, where only the model
  coefficients are calibrated. }

\subsection{Limitation of the Data Driven Approach}

In the proposed inverse modeling approach, observation data are used to infer the optimal
representation of complex structures in CFD simulations. The optimization is based on the how well a
representation can allow the numerical simulations to reproduce the observation data. The procedure
implicitly assumes that the numerical modeled system is the same as the observed system. However, in
reality this may not be true, since errors can come from many sources other than the parameterized
representation of the complex structure. They include numerical discretization error, other model
errors (e.g., those due to turbulence models), the discrepancies between the numerical simulation
setup and the experiments (e.g., the lack of representation of water tunnel side walls). 
When the procedure is used in practical applications, it is possible that
the inferred parameterization may be different from the most physical one. It is because the compensation 
for other errors that caused discrepancies between the simulated and observed responses. This error
compensation may be acceptable or even desirable for operational forecasting, since it is not
essential to find the most physical representation of the structure. Rather, the objective is to
obtain the best prediction of the unobserved quantities and regions based the observation and the
numerical model. However, when the approach is used obtain physically faithful representation of the
complex structure and use it for other simulations, caution must be exercised to minimize the
aforementioned discrepancies, and the remaining uncertainties should be parameterized and inferred
as well.

\section{Conclusion}
\label{sec:con} 
Simulating the flow through complex structures is challenging due to the prohibitive computational 
cost of the first-principle models and the unsatisfactory fidelities of the parameterized models.
In this work we proposed a data-driven, physics-based inverse modeling approach to improve the model
predictions for the flow with complexed structures by incorporating sparse measurement data.  The
effects of complex structures are represented by a non-parametric spatial force distribution, which
is inferred based on an iterative ensemble based Kalman method. The flow past a porous disk was
studied as an example problem to demonstrate the merits of the proposed inverse modeling scheme. A test
case with synthetic observation data is used to verify the proposed method, and the results show
that the inferred force distribution agrees well with the synthetic truth. A laboratory experiment
is conducted and the measurement data are used to perform a realistic inversion case. The simulation
results with the inferred force distribution are compared to those with the uniformly and
parabolically distributed forces and are validated by the experimental data. The comparisons
indicate that by using the proposed scheme the simulation results are markedly improved,
demonstrating a satisfactory performance of the inverse modeling approach on realistic
applications. The proposed data-driven inverse modeling approach is a promising tool to simulate the
flows through complex structures.

\appendix

\section{Iterative Ensemble Kalman Method for Inverse Modeling}
\label{app:enkf}

The algorithm of the iterative ensemble Kalman method for inverse modeling is summarized
below. See~\cite{iglesias2013ensemble} for details. Given the prior of the force distributions,
the follow steps are performed.
\begin{enumerate}
\item \textbf{(Sampling step)} Generate initial ensemble $\{{\bs{x}_j}\}_{j = 1}^{N}$ of size $N$,
  where the augmented system state is:
  \begin{equation}
    \label{eq:ini-x}
    \bs{x}_j  = [\bs{u}, \bs{k}, \bs{\omega}]_j  
    \notag
  \end{equation}

\item \textbf{(Prediction step)} 
  \begin{enumerate}
  \item Propagate the state from current state $n$ to the next iteration level $n+1$ by solving the 
  RANS equations (Eq.~\ref{eq:common-u}), indicated as $\mathcal{F}$,
  \begin{equation}
    \label{eq:forward}
    \hat{\bs{x}}_j^{(n+1)} = \mathcal{F} [ \bs{x}_j^{(n)} ]
    \notag
  \end{equation}
  This step involves reconstructing drag force fields for each sample and computing the
  velocities and turbulence kinetic energy from the RANS equations.
\item
  Estimate the mean $\bar{\bs{x}}$ and covariance $P^{(n+1)}$ of the ensemble as:
  \begin{subequations}
    \begin{align}
      &\bar{\bs{x}}^{(n+1)} = \frac{1}{N}\sum_{j=1}^{N}{\hat{\bs{x}}^{(n+1)}_j}     \notag  \\ 
      &P^{(n+1)} = \frac{1}{{N}-1} \sum_{j = 1}^{N} {\left( \hat{\bs{x}}_j\hat{\bs{x}}_j^T -  
          \bar{\bs{x}}\bar{\bs{x}}^T \right)^{(n+1)}} \notag   
    \end{align}
  \end{subequations}
\end{enumerate}

\item  \textbf{(Correction step)}
  \begin{enumerate}
  \item Compute the Kalman gain matrix as:
    \begin{equation}
      \label{eq:kalman-gain}
      K^{(n+1)} = P^{(n+1)} H^T (H P^{(n+1)} H^T + R)^{-1},
      \notag
    \end{equation}
where $H$ is the observation matrix, which project the full states to the observed
state. 
  \item Update each sample in the predicted ensemble as follows:
        \begin{equation}
      \label{eq:update}
      \bs{x}_j^{(n+1)} = \hat{\bs{x}}_j^{(n+1)} + K (\bs{y}_o - H \hat{\bs{x}}_j^{(n+1)}) 
      \notag
    \end{equation}
  \end{enumerate}

\item Repeat the prediction and correction steps until  the ensemble is statistically converged.
\end{enumerate}


\section{Notation}

\begin{tabbing}
  0000000\= this is definition\kill 
  $c$ \> degree of Jacobi polynomial \\
  $D$ \> diameter of the disk \\
  $f$	\> thrust force \\
  $f_t$	\> thrust force magnitude\\
  $H$ \> observation matrix \\
  $k$ \> turbulence kinetic energy \\
  $K$ \> Kalman gain matrix \\
  $\bs{k}$ \> turbulence kinetic energy state\\  
  $n$ \> number of mesh grids \\ 
  $N$ \> number of basis functions used \\
  $M$ \> number of samples \\
  $p$ \> pressure \\
  $P$ \> ensemble covariance\\
  $r$ \> polar coordinate \\
  $R$ \> observation error covariance \\ 
  $s$ \> argument of Jacobi polynomial $s = 2r^2 -1$ \\
  $t$ \> time coordinate \\
  $\bs{u}$ \> velocity state \\
  $U$ \> mean velocity \\
  $V$ \> volume of the computational domain \\
  $x$ \> spatial coordinate \\
  $\bs{x}$ \> state vector \\
  $\bs{y}_o$ \> observation data \\ 

{Greek letters}{}\\
$\rho$ \> density \\
$\nu$ \> viscosity \\
$\tau$ \> Reynolds stress \\
$\epsilon$ \> dissipation rate \\
$\phi$ \> orthogonal basis function \\
$\omega$ \> coefficients for force distribution \\
$\bs{\omega}$ \> coefficients vector \\
$\Omega$ \> computational domain \\
$\mathbf{\Omega}$ \> normalized force distribution function \\
$\theta$ \> polar coordinate \\
$\delta$ \> Kronecker delta

{Subscripts/Superscripts}{}\\

{Decorative symbols}{}\\
$\tilde{\Box}$ \> synthetic truth \\
$\bar{\Box}$ \> mean \\
$\hat{\Box}$ \> propagated state before correction  \\
$\Box '$ \> vector/matrix transpose \\
$\Box_o$ \> observed quantity
\end{tabbing}

\clearpage


\end{document}